\documentclass[showkeys, showpacs]{revtex4}
\usepackage{indentfirst}
\begin{document}
\title{\bf Baryonium $X(1835)$}
\author{S.M. Gerasyuta}
\email{gerasyuta@SG6488.spb.edu}
\author{E.E. Matskevich}
\email{matskev@pobox.spbu.ru}
\affiliation{Department of Theoretical Physics, St. Petersburg State University, 198904,
St. Petersburg, Russia}
\affiliation{Department of Physics, LTA, 194021, St. Petersburg, Russia}
\begin{abstract}
The relativistic six-quark equations including the $u$, $d$ quarks and
antiquarks are found. The nonstrange baryonia $B \bar B$ are constructed
without the mixing of the quarks and antiquarks. The relativistic six-quark
amplitudes of the baryonia are calculated. The poles of these amplitudes
determine the masses of baryonia. 16 masses of baryonia are predicted.
\end{abstract}
\maketitle
\section{Introduction.}
BES Collaboration observed a significant threshold enhancement of $p\bar p$
mass spectrum in the radiative decay $J/\psi\to\gamma\, p\bar p$ \cite{1}.
Recently BES Collaboration reported the results on $X(1835)$
in the $J/\psi\to\gamma\, \eta'\pi^+ \pi^-$ channel \cite{2, 3}. Under the strong
assumption that the $p\bar p$ threshold enhancement and $X(1835)$ are the
same resonance, Zhu and Gao suggested $X(1835)$ could be a $J^{PC}=0^{-+}$
$I^G=0^+$ $p\bar p$ baryonium \cite{4}.

Theoretical work speculated many possibilities for the enhancement
such as the t-channel pion exchange, some kind of threshold kinematical
effects, as new resonance below threshold or $p\bar p$ bound state
\cite{5, 6, 7, 8, 9, 10, 11, 12}.

In a series of papers \cite{13, 14, 15, 16, 17} a method has been developed which is
convenient for analysing relativistic three-hadron systems. The physics of
the three-hadron system can be described by means of a pair interaction
between the particles. There are three isobar channels, each of which
consists of a two-particle isobar and the third particle. The presence
of the isobar representation together with the condition of unitarity in
the pair energies and of analyticity leads to a system of integral equations
in a single variable. Their solution makes it possible to describe the
interaction of the produced particles in three-hadron systems.

In our papers \cite{18, 19, 20} relativistic generalization of the three-body
Faddeev equations was obtained in the form of dispersion relations in the
pair energy of two interacting quarks. The mass spectrum of $S$-wave
baryons including $u$, $d$, $s$ quarks was calculated by a method based on
isolating the leading singularities in the amplitude. We searched for the
approximate solution of integral three-quark equations by taking into
account two-particle and triangle singularities, all the weaker ones being
neglected. If we considered such approximation, which corresponds to
taking into account two-body and triangle singularities, and defined all
the smooth functions of the subenergy variables (as compared with the
singular part of the amplitude) in the middle point of the physical region
of Dalitz-plot, then the problem was reduced to the one of solving a system
of simple algebraic equations.

In the previous paper \cite{21} the relativistic six-quark equations are found in
the framework of coupled-channel formalism. The dynamical mixing between
the subamplitudes of hexaquark are considered. The six-quark amplitudes
of dibaryons are calculated. The poles of these amplitudes determine the
masses of dibaryons. We calculated the contribution of six-quark
subamplitudes to the hexaquark amplitudes.

In the present paper the relativistic six-quark equations including $u$, $d$
quarks and antiquarks are found. The nonstrange barionia $B \bar B$ are
constructed without the mixing of the quarks and antiquarks. The
relativistic six-quark amplitudes of the baryonia are calculated.
The poles of these amplitudes determine the masses of baryonia. In Sec. II the
six-quark amplitudes of baryonia are constructed. The dynamical mixing
between the subamplitudes of baryonia is considered. The relativistic
six-quark equations are obtained in the form of the dispersion relations
over the two-body subenergy. The approximate solutions of these equations
using the method based on the extraction of leading singularities of the
amplitude are obtained. Sec. III is devoted to the calculation results for
the baryonia mass spectrum and the contributions of subamplitudes to the
baryonia amplitude (Tables I, II, III, IV). In conclusion, the status
of the considered model is discussed.

\section{Six-quark amplitudes of the baryonia.}
As explained in the previous paper \cite{21} the relativistic
generalization of the three-body Faddeev equations was obtained in the
form of dispersion relations in the pair energy of two interacting quarks.
The pair quarks amplitudes $qq\rightarrow qq$ are calculated in
the framework of the dispersion $N/D$ method with the input four-fermion
interaction \cite{22, 23} with quantum numbers of the gluon \cite{24, 25}.

The construction of the approximate solution is based on
extraction of the leading singularities are close to the region
$s_{ik}\approx 4m^2$. Such a classification of singularities
makes it possible to search for an approximate solution of equations,
taking into account a definite number of leading singularities and
neglecting the weaker ones.

We derive the relativistic six-quark equations in the framework of the
dispersion relation technique. We use only planar diagrams; the other
diagrams due to the rules of $1/N_c$ expansion \cite{26, 27, 28} are neglected.
The current generates a six-quark system. The correct equations for the
amplitude are obtained by taking into account all possible subamplitudes.
It corresponds to the division of complete system into subsystems with a
smaller number of particles. Then one should represent a six-particle
amplitude as a sum of 15 subamplitudes:

\begin{equation}
A=\sum\limits_{i<j \atop i, j=1}^6 A_{ij}\, . \end{equation}

This defines the division of the diagrams into groups according to the
certain pair interaction of particles. The total amplitude can be
represented graphically as a sum of diagrams. We need to consider only
one group of diagrams and the amplitude corresponding to them, for example
$A_{12}$. We shall consider the derivation of the relativistic
generalization of the Faddeev-Yakubovsky approach.

In our case the low-lying baryonia are considered. We take into account the
pairwise interaction of all quarks and antiquarks in the baryonia.

The system of graphical equations Fig. 1 is determined using the
selfconsistent method. The coefficients are determined by the permutation
of quarks \cite{29, 30}. We should discuss the coefficient multiplying of the
diagrams in the equations of Fig. 1. For example, we consider the first
subamplitude $A_1^{1^{uu}}(s,s_{12345},s_{1234},s_{123},s_{12})$. In the
Fig. 1 the first coefficient is equal to 2 (permutation particles 1 and 2).
The second coefficient is equal to $6=2$ (permutation particles 1 and 2)
$\times 3$ (we consider the third, the fifth, the sixth particles). The
similar approach allows us to take into account the coefficients in all
equations.

In order to represent the subamplitudes $A_1^{1^{uu}}$, $A_1^{1^{u\bar d}}$,
$A_1^{1^{\bar d \bar d}}$, $A_2^{1^{uu}1^{\bar d \bar d}}$,
$A_3^{1^{uu}1^{u\bar d}1^{\bar d \bar d}}$ in the form of dispersion
relations, it is necessary to define the amplitudes of $qq$ and $q\bar q$
interactions.

We use the results of our relativistic quark model \cite{25} and write down the
pair quark amplitudes in the form:

\begin{equation}
a_n(s_{ik})=\frac{G^2_n(s_{ik})}
{1-B_n(s_{ik})} \, ,\end{equation}

\begin{equation}
B_n(s_{ik})=\int\limits_{(m_i+m_k)^2}^{\frac{(m_i+m_k)^2\Lambda}{4}}
\hskip2mm \frac{ds'_{ik}}{\pi}\frac{\rho_n(s'_{ik})G^2_n(s'_{ik})}
{s'_{ik}-s_{ik}} \, .\end{equation}

In the case of the $s$-channel amplitudes we use the matrix element:
$\left(\bar q_c^a O^i q^{a'} \right) \left(\bar q^{b'} O^i q_c^b \right)$,
where $q_c=\bar q C$ is the charge-conjugated spinor. $O^i$ are operators of different types
of the four-fermion interaction $(i=S, V, T, A, P)$, $a$, $b$, $a'$, $b'$ flavor indices.

Here $G_n(s_{ik})$ are the diquark vertex functions (Table V). The vertex
functions are determined by the contribution of the crossing channels.
The vertex functions satisfy the Fierz relations. Since the vertex functions depend only slightly
on energy it is possible to treat them as constants in our approximation and determine $G^2=N$
of $N/D$ method. These vertex functions are generated from gluon coupling constant $g$.
$B_n(s_{ik})$ and $\rho_n (s_{ik})$ are the Chew-Mandelstam functions with
cutoff $\Lambda$ \cite{31} and the phase spaces, respectively:

\begin{eqnarray}
\rho_n (s_{ik},J^{PC})&=&\left(\alpha(n,J^{PC}) \frac{s_{ik}}{(m_i+m_k)^2}
+\beta(n,J^{PC})+\delta(n,J^{PC}) \frac{(m_i-m_k)^2}{s_{ik}}\right)
\nonumber\\
&&\nonumber\\
&\times & \frac{\sqrt{(s_{ik}-(m_i+m_k)^2)(s_{ik}-(m_i-m_k)^2)}}
{s_{ik}}\, .
\end{eqnarray}

The coefficients $\alpha(n,J^{PC})$, $\beta(n,J^{PC})$ and
$\delta(n,J^{PC})$ are given in Table V.

Here $n=1$ coresponds to $q\bar q$-pairs with $J^P=0^-$, $n=2$ corresponds
to the $q\bar q$-pairs with $J^P=1^-$, $n=3$ defines the $qq$-pairs with
$J^P=0^+$, $n=4$ coresponds to $J^P=1^+$ $qq$-pairs.

Let us extract two- and three-particle singularities in the amplitudes
$A_1^{1^{uu}}(s,s_{12345},s_{1234},s_{123},s_{12})$,
$A_1^{1^{\bar d \bar d}}(s,s_{12345},s_{1234},s_{123},s_{12})$,
$A_1^{1^{u \bar d}}(s,s_{12345},s_{1234},s_{123},s_{12})$,
$A_2^{1^{uu}1^{\bar d \bar d}}(s,s_{12345},s_{1234},s_{12},s_{34})$,\\
$A_3^{1^{uu}1^{u \bar d}1^{\bar d \bar d}}
(s,s_{12345},s_{12},s_{34},s_{56})$:

\begin{eqnarray}
A_1^{1^{uu}}(s,s_{12345},s_{1234},s_{123},s_{12})&=
&\frac{\alpha_1^{1^{uu}} (s,s_{12345},s_{1234},s_{123},s_{12})
B_{1^{uu}}(s_{12})}{[1-B_{1^{uu}}(s_{12})]} \, ,\\
&&\nonumber\\
A_1^{1^{\bar d \bar d}}(s,s_{12345},s_{1234},s_{123},s_{12})&=
&\frac{\alpha_1^{1^{\bar d \bar d}} (s,s_{12345},s_{1234},s_{123},s_{12})
B_{1^{\bar d \bar d}}(s_{12})}{[1-B_{1^{\bar d \bar d}}(s_{12})]} \, ,\\
&&\nonumber\\
A_1^{1^{u \bar d}}(s,s_{12345},s_{1234},s_{123},s_{12})&=
&\frac{\alpha_1^{1^{u \bar d}} (s,s_{12345},s_{1234},s_{123},s_{12})
B_{1^{u \bar d}}(s_{12})}{[1-B_{1^{u \bar d}}(s_{12})]} \, ,\\
&&\nonumber\\
A_2^{1^{uu}1^{\bar d \bar d}}(s,s_{12345},s_{1234},s_{12},s_{34})&=
&\frac{\alpha_2^{1^{uu}1^{\bar d \bar d}}
(s,s_{12345},s_{1234},s_{12},s_{34})
B_{1^{uu}}(s_{12})B_{1^{\bar d \bar d}}(s_{34})}{[1-B_{1^{uu}}(s_{12})]
[1-B_{1^{\bar d \bar d}}(s_{34})]} \, , \\
&&\nonumber\\
A_3^{1^{uu}1^{u \bar d}1^{\bar d \bar d}}(s,s_{12345},s_{12},s_{34},s_{56})
&=&\frac{\alpha_3^{1^{uu}1^{u \bar d}1^{\bar d \bar d}}
(s,s_{12345},s_{12},s_{34},s_{56})
B_{1^{uu}}(s_{12})B_{1^{u \bar d}}(s_{34}) B_{1^{\bar d \bar d}}(s_{56})}
{[1- B_{1^{uu}}(s_{12})] [1- B_{1^{u \bar d}}(s_{34})]
[1- B_{1^{\bar d \bar d}}(s_{56})]}
 \, . \nonumber\\
&&
\end{eqnarray}

We do not extract four-particles singularities, because they are weaker
than two- and three-particle singularities.

We used the classification of singularities, which was proposed in
paper \cite{32}. The construction of the approximate solution of Eqs.
(5) -- (9) is based on the extraction of the leading singularities
of the amplitudes. The main singularities in $s_{ik}=(m_i+m_k)^2$
are from pair rescattering of the particles $i$ and $k$. First of all there
are threshold square-root singularities. Also possible are pole
singularities which correspond to the bound states. The diagrams of Fig. 1
apart from two-particle singularities have triangular singularities and the
singularities defining the interactions of four, five and six particles.
Such classification allows us to search the corresponding solution of
equations by taking into account some definite number of leading
singularities and neglecting all the weaker ones. We consider the
approximation which defines two-particle, triangle and four-,  five- and
six-particle singularities. The contribution of two-particle and triangle
singularities are more important, but we must take into account also the
other singularities.

The five functions $\alpha_i$ are the smooth functions of $s_{ik}$,
$s_{ijk}$, $s_{ijkl}$ $s_{ijklm}$ as compared with the singular part of the
amplitudes, hence they can be expanded in a series in the singularity point
and only the first term of this series should be employed further. Using
this classification, one defines the reduced amplitudes $\alpha_i$ as well
as the $B$-functions in the middle point of physical region of Dalitz-plot
at the point $s_0$:

\begin{eqnarray}
s_0=\frac{s+4\sum\limits_{i=1}^{6} m_i^2}
{\sum\limits_{i,k=1 \atop i<k}^{6} m_{ik}^2}
\, ,
\end{eqnarray}

\begin{eqnarray}
s_{123}=s_0 \sum\limits_{i,k=1 \atop i<k}^{3} m_{ik}^2
-\sum\limits_{i=1}^{3} m_i^2
\, ,
\end{eqnarray}

\begin{eqnarray}
s_{1234}=s_0 \sum\limits_{i,k=1 \atop i<k}^{4} m_{ik}^2
-2\sum\limits_{i=1}^{4} m_i^2
\, .
\end{eqnarray}

Such choice of point $s_0$ allows us to replace integral equations
(Fig. 1) by the algebraic equations (13) -- (17), respectively:

\begin{eqnarray}
%1
\alpha_1^{1^{uu}} &=&\lambda+2 I_1(1^{uu}1^{uu}) \alpha_1^{1^{uu}}
+6 I_1(1^{uu}1^{u \bar d}) \alpha_1^{1^{u \bar d}}
\, , \\
&&\nonumber\\
%2
\alpha_1^{1^{\bar d \bar d}} &=&\lambda
+2 I_1(1^{\bar d \bar d}1^{\bar d \bar d}) \alpha_1^{1^{\bar d \bar d}}
+6 I_1(1^{\bar d \bar d}1^{u \bar d}) \alpha_1^{1^{u \bar d}}
\, , \\
&&\nonumber\\
%3
\alpha_1^{1^{u \bar d}} &=&\lambda
+2 I_1(1^{u \bar d}1^{uu}) \alpha_1^{1^{uu}}
+2 I_1(1^{u \bar d}1^{\bar d \bar d}) \alpha_1^{1^{\bar d \bar d}}
+4 I_1(1^{u \bar d}1^{u \bar d}) \alpha_1^{1^{u \bar d}}
+4 I_2(1^{u \bar d}1^{uu}1^{\bar d \bar d})
\alpha_2^{1^{uu}1^{\bar d \bar d}}
\, , \nonumber\\
&&\\
%4
\alpha_2^{1^{uu}1^{\bar d \bar d}} &=&\lambda
+2 I_4(1^{uu}1^{\bar d \bar d}1^{uu}) \alpha_1^{1^{uu}}
+2 I_4(1^{\bar d \bar d}1^{uu}1^{\bar d \bar d})
\alpha_1^{1^{\bar d \bar d}}
+4 I_3(1^{uu}1^{\bar d \bar d}1^{u\bar d}) \alpha_1^{1^{u \bar d}}
\nonumber\\
&&\nonumber\\
&+&4 I_6(1^{uu}1^{\bar d \bar d}1^{uu}1^{\bar d \bar d})
\alpha_2^{1^{uu}1^{\bar d \bar d}}
+4 I_8(1^{uu}1^{\bar d \bar d}1^{uu}1^{u \bar d}1^{\bar d \bar d})
\alpha_3^{1^{uu}1^{u \bar d}1^{\bar d \bar d}}
\, , \nonumber\\
&& \\
%5
\alpha_3^{1^{uu}1^{u \bar d}1^{\bar d \bar d}}&=&\lambda
+2 I_9(1^{uu}1^{u \bar d}1^{\bar d \bar d}1^{uu}) \alpha_1^{1^{uu}}
+2 I_9(1^{\bar d \bar d}1^{u \bar d}1^{uu}1^{\bar d \bar d})
\alpha_1^{1^{\bar d \bar d}}
+2 I_9(1^{uu}1^{u \bar d}1^{\bar d \bar d}1^{u \bar d})
\alpha_1^{1^{u \bar d}}
\nonumber\\
&&\nonumber\\
&+&4 I_9(1^{uu}1^{\bar d \bar d}1^{u \bar d}1^{u \bar d})
\alpha_1^{1^{u \bar d}}
+2 I_9(1^{\bar d \bar d}1^{u \bar d}1^{uu}1^{u \bar d})
\alpha_1^{1^{u \bar d}}
+4 I_{10}(1^{uu}1^{u \bar d}1^{\bar d \bar d}1^{uu}1^{\bar d \bar d})
\alpha_2^{1^{uu}1^{\bar d \bar d}} \, , \nonumber\\
&&
\end{eqnarray}

\noindent
where $\lambda_i$ are the current constants. We used the functions
$I_1$, $I_2$, $I_3$, $I_4$, $I_6$, $I_8$, $I_9$, $I_{10}$:

\begin{eqnarray}
I_1(ij)&=&\frac{B_j(s_0^{13})}{B_i(s_0^{12})}
\int\limits_{(m_1+m_2)^2}^{\frac{(m_1+m_2)^2\Lambda_i}{4}}
\frac{ds'_{12}}{\pi}\frac{G_i^2(s_0^{12})\rho_i(s'_{12})}
{s'_{12}-s_0^{12}} \int\limits_{-1}^{+1} \frac{dz_1(1)}{2}
\frac{1}{1-B_j (s'_{13})}\, , \\
&&\nonumber\\
I_2(ijk)&=&\frac{B_j(s_0^{13}) B_k(s_0^{24})}{B_i(s_0^{12})}
\int\limits_{(m_1+m_2)^2}^{\frac{(m_1+m_2)^2\Lambda_i}{4}}
\frac{ds'_{12}}{\pi}\frac{G_i^2(s_0^{12})\rho_i(s'_{12})}
{s'_{12}-s_0^{12}}
\frac{1}{2\pi}\int\limits_{-1}^{+1}\frac{dz_1(2)}{2}
\int\limits_{-1}^{+1} \frac{dz_2(2)}{2}\nonumber\\
&&\nonumber\\
&\times&
\int\limits_{z_3(2)^-}^{z_3(2)^+} dz_3(2)
\frac{1}{\sqrt{1-z_1^2(2)-z_2^2(2)-z_3^2(2)+2z_1(2) z_2(2) z_3(2)}}
\nonumber\\
&&\nonumber\\
&\times& \frac{1}{1-B_j (s'_{13})} \frac{1}{1-B_k (s'_{24})}
 \, , \\
&&\nonumber\\
I_3(ijk)&=&\frac{B_k(s_0^{23})}{B_i(s_0^{12}) B_j(s_0^{34})}
\int\limits_{(m_1+m_2)^2}^{\frac{(m_1+m_2)^2\Lambda_i}{4}}
\frac{ds'_{12}}{\pi}\frac{G_i^2(s_0^{12})\rho_i(s'_{12})}
{s'_{12}-s_0^{12}}\nonumber\\
&&\nonumber\\
&\times&\int\limits_{(m_3+m_4)^2}^{\frac{(m_3+m_4)^2\Lambda_j}{4}}
\frac{ds'_{34}}{\pi}\frac{G_j^2(s_0^{34})\rho_j(s'_{34})}
{s'_{34}-s_0^{34}}
\int\limits_{-1}^{+1} \frac{dz_1(3)}{2} \int\limits_{-1}^{+1}
\frac{dz_2(3)}{2} \frac{1}{1-B_k (s'_{23})} \, , \\
&&\nonumber\\
I_4(ijk)&=&I_1(ik) \, , \\
&&\nonumber\\
I_6(ijkl)&=&I_1(ik) \cdot I_1(jl)
 \, , \\
&&\nonumber\\
I_8(ijklm)&=&\frac{B_k(s_0^{15})B_l(s_0^{23})B_m(s_0^{46})}
{B_i(s_0^{12}) B_j(s_0^{34})}
\int\limits_{(m_1+m_2)^2}^{\frac{(m_1+m_2)^2\Lambda_i}{4}}
\frac{ds'_{12}}{\pi}\frac{G_i^2(s_0^{12})\rho_i(s'_{12})}
{s'_{12}-s_0^{12}}\nonumber\\
&&\nonumber\\
&\times&\int\limits_{(m_3+m_4)^2}^{\frac{(m_3+m_4)^2\Lambda_j}{4}}
\frac{ds'_{34}}{\pi}\frac{G_j^2(s_0^{34})\rho_j(s'_{34})}
{s'_{34}-s_0^{34}}\nonumber\\
&&\nonumber\\
&\times&\frac{1}{(2\pi)^2}\int\limits_{-1}^{+1}\frac{dz_1(8)}{2}
\int\limits_{-1}^{+1} \frac{dz_2(8)}{2}
\int\limits_{-1}^{+1} \frac{dz_3(8)}{2}
\int\limits_{z_4(8)^-}^{z_4(8)^+} dz_4(8)
\int\limits_{-1}^{+1} \frac{dz_5(8)}{2}
\int\limits_{z_6(8)^-}^{z_6(8)^+} dz_6(8)
\nonumber\\
&&\nonumber\\
&\times&
\frac{1}{\sqrt{1-z_1^2(8)-z_3^2(8)-z_4^2(8)+2z_1(8) z_3(8) z_4(8)}}
\nonumber\\
&&\nonumber\\
&\times&
\frac{1}{\sqrt{1-z_2^2(8)-z_5^2(8)-z_6^2(8)+2z_2(8) z_5(8) z_6(8)}}
\nonumber\\
&&\nonumber\\
&\times& \frac{1}{1-B_k (s'_{15})} \frac{1}{1-B_l (s'_{23})}
\frac{1}{1-B_m (s'_{46})}
 \, , \\
&&\nonumber\\
I_9(ijkl)&=&I_3(ijl) \, , \\
&&\nonumber\\
I_{10}(ijklm)&=
&\frac{B_l(s_0^{23})B_m(s_0^{45})}
{B_i(s_0^{12}) B_j(s_0^{34}) B_k(s_0^{56})}
\int\limits_{(m_1+m_2)^2}^{\frac{(m_1+m_2)^2\Lambda_i}{4}}
\frac{ds'_{12}}{\pi}\frac{G_i^2(s_0^{12})\rho_i(s'_{12})}{s'_{12}-s_0^{12}}
\nonumber\\
&&\nonumber\\
&\times&
\int\limits_{(m_3+m_4)^2}^{\frac{(m_3+m_4)^2\Lambda_j}{4}}
\frac{ds'_{34}}{\pi}\frac{G_j^2(s_0^{34})\rho_j(s'_{34})}
{s'_{34}-s_0^{34}}
\int\limits_{(m_5+m_6)^2}^{\frac{(m_5+m_6)^2\Lambda_k}{4}}
\frac{ds'_{56}}{\pi}\frac{G_k^2(s_0^{56})\rho_k(s'_{56})}{s'_{56}-s_0^{56}}
\nonumber\\
&&\nonumber\\
&\times&
\frac{1}{2\pi}\int\limits_{-1}^{+1}\frac{dz_1(10)}{2}
\int\limits_{-1}^{+1} \frac{dz_2(10)}{2}
\int\limits_{-1}^{+1} \frac{dz_3(10)}{2}
\int\limits_{-1}^{+1} \frac{dz_4(10)}{2}
\int\limits_{z_5(1-)^-}^{z_5(10)^+} dz_5(10)
\nonumber\\
&&\nonumber\\
&\times&
\frac{1}{\sqrt{1-z_1^2(10)-z_4^2(10)-z_5^2(10)+2z_1(10) z_4(10) z_5(10)}}
\nonumber\\
&&\nonumber\\
&\times& \frac{1}{1-B_l (s'_{23})} \frac{1}{1-B_m (s'_{45})}
 \, ,
\end{eqnarray}

\noindent
where $i$, $j$, $k$, $l$, $m$ correspond to the diquarks with the
spin-parity $J^P=0^+, 1^+$ and mesons with the spin-parity
$J^P=0^-, 1^-$.

The other choices of point $s_0$ do not change essentially the contributions
of $\alpha_i$, therefore we omit the indices $s_0^{ik}$. Since the
vertex functions depend only slightly on energy, it is possible to treat
them as constants in our approximation.

We can pass from the integration over cosines of the angles to the integration
over the subenergies \cite{21}. In the relativistic invariant solution the
center of mass of particles 1, 2 by the standard method is treated \cite{21}.

The system of graphical equations Fig. 1 is determined by the subamplitudes $A_1$, $A_2$, $A_3$.
$B \bar B$ states ($A_2$) are constructed without the mixing of quarks and antiquarks. Therefore
we did not use the three mesons and meson plus tetraquark states. But the subamplitudes $A_1$
and $A_3$ contain the quark-antiquark pairs. Then the algebraic equations (13) -- (17) take
into account the contributions reduced amplitudes $\alpha_1^{1^{uu}}$, $\alpha_1^{1^{\bar d \bar d}}$,
$\alpha_1^{1^{u \bar d}}$, $\alpha_2^{1^{uu}1^{\bar d \bar d}}$,
$\alpha_3^{1^{uu}1^{u \bar d}1^{\bar d \bar d}}$. The solution of the system of equations are
considered as:

\begin{equation}
\alpha_i(s)=F_i(s,\lambda_i)/D(s) \, ,\end{equation}

\noindent
where zeros of $D(s)$ determinants define the masses of bound states of baryonia.

We have analyzed in the subamplitudes of a quark and an antiquark and did not obtain the
bound state with the model parameters.

As example, we consider the equations for the quark content
$uuu \bar d \bar d \bar d$ with the isospin $I=3$ and the spin-parity
$J^P=3^-$ (Fig. 1). The similar equations have been calculated for the
isospin $I=0,\, 1, \, 2,\, 3$ and the spin-parity $J^P=0^-, 1^-, 2^-, 3^-$.
We take into account the $u$ and $d$ quarks.

The functions $I_1$, $I_2$, $I_3$, $I_4$, $I_6$, $I_8$, $I_9$, $I_{10}$
determine the interaction of the quarks and the antiquarks. These functions
take into account the contributions of the Chew-Mandelstam functions, which
are constructed in the model for the quark-antiquark pairs with various
quantum numders using the unitarity condition.

\section{Calculation results.}
The poles of the reduced amplitudes $\alpha_i$ correspond to the bound states
and determine the masses of the baryonia. The dynamical mixing between the
subamplitudes of baryonia is considered. We derived the relativistic six-quark
equations in the framework of the dispersion technique. The pair quarks
amplitudes $qq\rightarrow qq$ and $q \bar q\rightarrow q \bar q$ are calculated
with the dispersion $N/D$ method using the input four-fermion interaction
\cite{22, 23} with the quantum numbers of the gluon \cite{24, 25}.

The model under consideration proceeds from the assumption that the confinement radius is sufficiently
larger than constituent quark radii as well as the radii of the forces which bound the low-lying hadrons.
It means that quark interaction forces are the two component ones. The long-range component is due to the
confinement. In the present paper, when the low-lying hadrons are considered, the long-range component
of the forces is neglected.

We manage with the quarks as with real particles. However, in the soft region, the quark diagrams
should be treated as spectral integrals over quark masses with the spectral density $\rho(m^2)$:
the integration over quark masses in the amplitudes puts away the quark singularities and introduces
the hadron ones. We can believe that the approximation $\rho(m^2) \to \delta(m^2-m_q^2)$ could be possible
for the low-lying hadrons. We hope that this approach is sufficiently good for the calculation of the
low-lying baryonia being carried out here. The problem of the distribution over quark masses is important
when one considers the high-excited states.

The four-quark interaction is considered as an input:

\begin{equation}
g_V \left(\bar q \lambda I_f \gamma_{\mu} q \right)^2
\, . \end{equation}

\noindent
Here $I_f$ is the unity matrix in the flavor space $(u, d)$. $\lambda$ are
the color Gell-Mann matrices.

We introduce the scale of the dimensional parameters \cite{25}:

\begin{equation}
g=\frac{m^2}{\pi^2}g_V  \, , \quad
\Lambda=\frac{4\Lambda(ik)} {(m_i+m_k)^2}.
\end{equation}

\noindent
Here $m_i$ and $m_k$ are the quark masses in the intermediate state of
the quark loop. Dimensionless parameters $g$ and $\Lambda$ are supposed
to be the constants which are independent of the quark interaction type.
In the case under question the interacting pairs of particles do not
form the bound states. The attraction of quark-antiquark and quark-quark
pairs is not enough for the construction of the bound state. This is similar
to the case of the four-quark systems \cite{32}. Therefore, the integration in
the dispersion integral run from $4m^2$ to $\Lambda$.

The quark masses of model $m_{u, d}=410\, MeV$ coincide with ordinary
baryon ones \cite{21}. The model in question has only two parameters:
the cutoff parameter $\Lambda=11$ and the gluon coupling constant $g=0.314$.
These parameters are similar to the previous paper \cite{21} ones.

The estimation of theoretical error on the baryonia masses is $1\, MeV$.
This result was obtained by the choice of model parameters.

Some authors had investigated the tetraquark system with the three-body $qq \bar q$ and
$q \bar q \bar q$ interaction, whose existence has no direct effect on the ordinary hadron states
\cite{33, 34}. But assuming the most general two-body quark Hamiltonian \cite{35, 36}, Pirjol and Schat
derive universal correlations among masses and mixing angles which are valid in any model for
quark interactions containing only two-body interactions. Deviation from these predictions
provide no evidence for the presence of spin-flavor dependent three-body quark interactions.

Our model is based on the three principles of unitarity, analyticity and crossing symmetry. The
principle of unitarity are applied to the two-body subenergy channels.

Further experimental and theoretical efforts are required in order to satisfactory explain the
presence of three-body quark interactions.

In the Table I the calculated masses of nonstrange baryonia are shown.
The contributions of subamplitudes to the six-quark amplitude are
represented in the Tables II, III, IV. The states
($\Delta \bar \Delta+\Delta \bar n+n \bar \Delta+n\bar n$),
($\Delta \bar \Delta+\Delta \bar p+p \bar \Delta+p\bar p$) and
($\Delta \bar \Delta+\Delta \bar n+p \bar \Delta+p\bar n$) with the
isospins $I=0$, $1$ and the spin-parity $J^P=0^-$ possess the mass
$M=1835\, MeV$. We predict the degeneracy of the some states.
For the ($\Delta \bar \Delta+p \bar \Delta$),
($\Delta \bar \Delta+\Delta \bar n$),
($\Delta \bar \Delta+n \bar \Delta$) and
($\Delta \bar \Delta+\Delta \bar p$) with the isospins $I=1$, $2$ and
the spin-parity $J^P=0^-$ the mass $M=1928\, MeV$ is obtained.
The low-lying state $uuu \bar d \bar d \bar d$ $(\Delta \bar \Delta)$
with the isospin $I=3$ and the spin-parity $J^P=1^-$ possesses the mass
$M=1783\, MeV$. The dynamical mixing between the five subamplitudes
(similar to the Fig. 1) is considered. Therefore the multiquark state will
be stable. The states $uuu \bar u \bar u \bar u$ $(\Delta \bar \Delta)$
and $ddd \bar d \bar d \bar d$ $(\Delta \bar \Delta)$ with the isospin
$I=0$ and the spin-parity $J^P=0^-$ have the mass $M=1973\, MeV$.

We predict the degeneracy of baryonia
$M(uud \bar d \bar d \bar d,\, I=2)=M(uuu \bar u \bar d \bar d,\, I=2)
=M(udd \bar d \bar d \bar d,\, I=1)=M(uuu \bar u \bar u \bar d,\, I=1)$.
For the states
$M(uud \bar u \bar d \bar d,\, I=1)=M(udd \bar u \bar d \bar d,\, I=0)
=M(uud \bar u \bar u \bar d,\, I=0)$ and
$M(uuu \bar u \bar u \bar u,\, I=0)=M(ddd \bar d \bar d \bar d,\, I=0)$
the degeneracy is also obtained.

A somewhat simple picture of baryonium is that of a deuteron-like $N\bar N$
bound state or resonance, benefiting from the attractive potential mediated
by the exchange of gluon \cite{21}. We consider the influence of the contributions
of quark-antiquark pairs.

Entem and Fernandez, describing scattering data and mass shifts of $p\bar p$
levels in a constituent quark model, assign the threshold enhancement to
final-state interaction \cite{37, 38}. Zou and Chiang find that final state
interaction makes an important contribution to the $p\bar p$ near threshold
enhancement \cite{39}.

BES collaboration measured the mass $X(1835)$ to be $M=1833.7\, MeV$ and its
width to be $\Gamma=67.7\, MeV$ \cite{2}.

The state with $M=1835\, MeV$ is considered as $p\bar p$ state \cite{4}
or the second radial excitation of $\eta'$ meson \cite{40}.

In our case this state have following content
$\Delta \bar \Delta+\Delta \bar p+p \bar \Delta+p\bar p$ with isospin
$I=0$ and spin-parity $J^P=0^-$.

We calculated the masses of baryonia with the isospin $I=0,\, 1,\, 2,\, 3$
and spin-parity $J^P=0^-,\, 1^-,\, 2^-,\, 3^-$ (Table I).

\begin{acknowledgments}
The authors would like to thank T. Barnes and C.-Y. Wong for useful
discussions. The work was carried with the support of the Russian Ministry
of Education (grant 2.1.1.68.26).
\end{acknowledgments}

\newpage

\newpage

\begin{table}
\caption{S-wave baryonia masses. Parameters of model: cutoff
$\Lambda=11.0$, gluon coupling constant $g=0.314$.
Quark masses $m_{u,d}=410\, MeV$.}
\begin{tabular}{|c|c|c|c|}
\hline
$I$ & Quark content (baryonia) & $J$ & Mass (MeV) \\[5pt]
\hline
$0$ &
\begin{tabular}{l}
$uuu \bar u \bar u \bar u$ ($\Delta \bar \Delta$),\\
$ddd \bar d \bar d \bar d$ ($\Delta \bar \Delta$)
\end{tabular}
&
\begin{tabular}{c}
$0$ \\
$1$ \\
$2$ \\
$3$
\end{tabular}
&
\begin{tabular}{c}
$1973$ \\
$1824$ \\
$1938$ \\
$2290$
\end{tabular}
\\
\hline
$0$; $1$
&
\begin{tabular}{c}
$udd \bar u \bar d \bar d$
($\Delta \bar \Delta+\Delta \bar n+n \bar \Delta+n\bar n$),\\
 $uud \bar u \bar u \bar d$
($\Delta \bar \Delta+\Delta \bar p+p \bar \Delta+p\bar p$);\\
$uud \bar u \bar d \bar d$
($\Delta \bar \Delta+\Delta \bar n+p \bar \Delta+p\bar n$)
\end{tabular}
&
\begin{tabular}{c}
$0$ \\
$1$ \\
$2$ \\
$3$
\end{tabular}
&
\begin{tabular}{c}
$1835$ \\
$1784$ \\
$1851$ \\
$2455$
\end{tabular}
\\
\hline
$1$; $2$
&
\begin{tabular}{l}
$udd \bar d \bar d \bar d$ ($\Delta \bar \Delta+n \bar \Delta$),\\
$uuu \bar u \bar u \bar d$ ($\Delta \bar \Delta+\Delta \bar p$);\\
$uud \bar d \bar d \bar d$ ($\Delta \bar \Delta+p \bar \Delta$),\\
$uuu \bar u \bar d \bar d$ ($\Delta \bar \Delta+\Delta \bar n$)
\end{tabular}
&
\begin{tabular}{c}
$0$ \\
$1$ \\
$2$ \\
$3$
\end{tabular}
&
\begin{tabular}{c}
$1928$ \\
$1770$ \\
$1857$ \\
$2395$
\end{tabular}
\\
\hline
$3$ &
\begin{tabular}{l}
$uuu \bar d \bar d \bar d$ ($\Delta \bar \Delta$)
\end{tabular}
&
\begin{tabular}{c}
$0$ \\
$1$ \\
$2$ \\
$3$
\end{tabular}
&
\begin{tabular}{c}
$2067$ \\
$1783$ \\
$1938$ \\
$2290$
\end{tabular}
\\
\hline
\end{tabular}
\end{table}

\begin{table}
\caption{$IJ=00$, $uud \bar u \bar u \bar d$ $(1835 \, MeV)$,
$\Lambda=11.0$, $g=0.314$.}
\begin{tabular}{|c|c|}
\hline
 Subamplitudes & Contributions, percent \\
\hline
$A_1^{1^{uu}}$ & $3.7$ \\
$A_1^{1^{\bar u \bar u}}$ & $3.7$ \\
$A_1^{1^{u \bar u}}$ & $9.0$ \\
$A_1^{1^{u \bar d}}$ & $7.8$ \\
$A_1^{1^{d \bar u}}$ & $7.8$ \\
$A_1^{1^{d \bar d}}$ & $6.7$ \\
$A_1^{0^{ud}}$ & $3.6$ \\
$A_1^{0^{\bar u \bar d}}$ & $3.6$ \\
$A_1^{0^{u \bar u}}$ & $7.0$ \\
$A_1^{0^{u \bar d}}$ & $6.8$ \\
$A_1^{0^{d \bar u}}$ & $6.8$ \\
$A_1^{0^{d \bar d}}$ & $6.6$ \\
$A_2^{1^{uu}1^{\bar u \bar u}}$ & $2.4$ \\
$A_2^{1^{uu}0^{\bar u \bar d}}$ & $2.0$ \\
$A_2^{1^{\bar u \bar u}0^{ud}}$ & $2.6$ \\
$A_2^{0^{ud}0^{\bar u \bar d}}$ & $2.8$ \\
$A_3^{1^{uu}0^{d \bar d}1^{\bar u \bar u}}$ & $4.0$ \\
$A_3^{1^{uu}1^{d \bar u}0^{\bar u \bar d}}$ & $4.5$ \\
$A_3^{1^{\bar u \bar u}1^{u \bar d}0^{ud}}$ & $4.5$ \\
$A_3^{0^{ud}0^{u \bar u}0^{\bar u \bar d}}$ & $4.3$ \\
\hline
$\sum A_1$ & $73.0$ \\
$\sum A_2$ & $9.8$ \\
$\sum A_3$ & $17.2$ \\
\hline
\end{tabular}
\end{table}

\begin{table}
\caption{$IJ=33$, $uuu \bar d \bar d \bar d$ $(2290 \, MeV)$,
$\Lambda=11.0$, $g=0.314$.}
\begin{tabular}{|c|c|}
\hline
 Subamplitudes & Contributions, percent \\
\hline
$A_1^{1^{uu}}$ & $9.9$ \\
$A_1^{1^{\bar d \bar d}}$ & $9.9$ \\
$A_1^{1^{u \bar d}}$ & $25.4$ \\
$A_2^{1^{uu}1^{\bar d \bar d}}$ & $14.5$ \\
$A_3^{1^{uu}1^{\bar d \bar d}1^{u \bar d}}$ & $40.3$ \\
\hline
$\sum A_1$ & $45.2$ \\
$\sum A_2$ & $14.5$ \\
$\sum A_3$ & $40.3$ \\
\hline
\end{tabular}
\end{table}

\begin{table}
\caption{$IJ=31$, $uuu \bar d \bar d \bar d$ $(1783 \, MeV)$,
$\Lambda=11.0$, $g=0.314$.}
\begin{tabular}{|c|c|}
\hline
 Subamplitudes & Contributions, percent \\
\hline
$A_1^{1^{uu}}$ & $14.5$ \\
$A_1^{1^{\bar d \bar d}}$ & $14.5$ \\
$A_1^{1^{u \bar d}}$ & $34.8$ \\
$A_1^{0^{u \bar d}}$ & $25.2$ \\
$A_2^{1^{uu}1^{\bar d \bar d}}$ & $11.1$ \\
\hline
$\sum A_1$ & $88.9$ \\
$\sum A_2$ & $11.1$ \\
\hline
\end{tabular}
\end{table}

\begin{table}
\caption{Vertex functions and Chew-Mandelstam coefficients.}
\begin{tabular}{|c|c|c|c|c|}
\hline
$i$ & $G_i^2(s_{kl})$ & $\alpha_i$ & $\beta_i$ & $\delta_i$ \\
\hline
& & & & \\
$0^+$ diquark & $\frac{4g}{3}-\frac{8gm_{kl}^2}{(3s_{kl})}$
& $\frac{1}{2}$ & $-\frac{1}{2}\frac{(m_k-m_l)^2}{(m_k+m_l)^2}$ & $0$ \\
& & & & \\
$1^+$ diquark & $\frac{2g}{3}$ & $\frac{1}{3}$
& $\frac{4m_k m_l}{3(m_k+m_l)^2}-\frac{1}{6}$
& $-\frac{1}{6}\frac{(m_k-m_l)^2}{(m_k+m_l)^2}$ \\
& & & & \\
$0^-$ meson & $\frac{8g}{3}-\frac{16gm_{kl}^2}{(3s_{kl})}$
& $\frac{1}{2}$ & $-\frac{1}{2}\frac{(m_k-m_l)^2}{(m_k+m_l)^2}$ & $0$ \\
& & & & \\
$1^-$ meson & $\frac{4g}{3}$ & $\frac{1}{3}$
& $\frac{4m_k m_l}{3(m_k+m_l)^2}-\frac{1}{6}$
& $-\frac{1}{6}\frac{(m_k-m_l)^2}{(m_k+m_l)^2}$ \\
& & & & \\
\hline
\end{tabular}
\end{table}

\suppressfloats

\newpage

%\vskip30pt
\begin{figure}
\begin{picture}(600,100)
%1
\put(0,45){\line(1,0){18}}
\put(0,47){\line(1,0){17.5}}
\put(0,49){\line(1,0){17}}
\put(0,51){\line(1,0){17}}
\put(0,53){\line(1,0){17.5}}
\put(0,55){\line(1,0){18}}
\put(30,50){\circle{25}}
\put(19,46){\line(1,1){15}}
\put(22,41){\line(1,1){17}}
\put(27.5,38.5){\line(1,1){14}}
\put(31,63){\vector(1,1){20}}
\put(37,60){\vector(1,1){20}}
\put(31,38){\vector(1,-1){20}}
\put(37,40){\vector(1,-1){20}}
\put(50,50){\circle{16}}
\put(58,50){\vector(3,2){22}}
\put(58,50){\vector(3,-2){22}}
\put(75,70){$u$}
\put(83,64){\small 1}
\put(75,23){$u$}
\put(83,32){\small 2}
\put(47,86){$\bar d$}
\put(38,82){\small 4}
\put(60,81){$u$}
\put(62,70){\small 3}
\put(39,10){$\bar d$}
\put(35,20){\small 6}
\put(60,13){$\bar d$}
\put(58,27){\small 5}
\put(43,47){\footnotesize $1^{uu}$}
\put(90,48){$=$}
%2
\put(110,45){\line(1,0){18}}
\put(110,47){\line(1,0){17.5}}
\put(110,49){\line(1,0){17}}
\put(110,51){\line(1,0){17}}
\put(110,53){\line(1,0){17.5}}
\put(110,55){\line(1,0){18}}
\put(135,50){\circle{16}}
\put(128,55){\vector(3,2){22}}
\put(128,55){\vector(2,3){15}}
\put(128,45){\vector(3,-2){22}}
\put(128,45){\vector(2,-3){15}}
\put(143,50){\vector(3,1){22}}
\put(143,50){\vector(3,-1){22}}
\put(167,60){$u$}
\put(160,60){\small 1}
\put(167,33){$u$}
\put(160,34){\small 2}
\put(152,64){$u$}
\put(152,72){\small 3}
\put(144,81){$\bar d$}
\put(136,80){\small 4}
\put(152,21){$\bar d$}
\put(151,33){\small 5}
\put(142,12){$\bar d$}
\put(134,16){\small 6}
\put(128,47){\footnotesize $1^{uu}$}
\put(175,48){$+$}
%3
\put(190,48){2}
\put(203,45){\line(1,0){18}}
\put(203,47){\line(1,0){17.5}}
\put(203,49){\line(1,0){17}}
\put(203,51){\line(1,0){17}}
\put(203,53){\line(1,0){17.5}}
\put(203,55){\line(1,0){18}}
\put(233,50){\circle{25}}
\put(222,46){\line(1,1){15}}
\put(225,41){\line(1,1){17}}
\put(230.5,38.5){\line(1,1){14}}
\put(249,62){\circle{16}}
\put(254.5,68.5){\vector(1,1){15}}
\put(254.5,68.5){\vector(1,-1){18}}
\put(245,50){\vector(1,0){28}}
\put(281,50){\circle{16}}
\put(289,50){\vector(3,1){22}}
\put(289,50){\vector(3,-1){22}}
\put(243,43){\vector(1,-1){15}}
\put(240.5,40.5){\vector(2,-3){12}}
\put(237.5,38){\vector(1,-3){6.5}}
\put(265,61){$u$}
\put(260,65){\small 1}
\put(260,40){$u$}
\put(253,40){\small 2}
\put(310,60){$u$}
\put(303,60){\small 1}
\put(310,33){$u$}
\put(302,33){\small 2}
\put(273,82){$u$}
\put(260,82){\small 3}
\put(264,21){$\bar d$}
\put(261,30){\small 4}
\put(253,10){$\bar d$}
\put(254,22){\small 5}
\put(242,7){$\bar d$}
\put(236,10){\small 6}
\put(274,47){\footnotesize $1^{uu}$}
\put(242,59){\footnotesize $1^{uu}$}
\put(320,48){$+$}
%4
\put(335,48){6}
\put(348,45){\line(1,0){18}}
\put(348,47){\line(1,0){17.5}}
\put(348,49){\line(1,0){17}}
\put(348,51){\line(1,0){17}}
\put(348,53){\line(1,0){17.5}}
\put(348,55){\line(1,0){18}}
\put(378,50){\circle{25}}
\put(367,46){\line(1,1){15}}
\put(370,41){\line(1,1){17}}
\put(375.5,38.5){\line(1,1){14}}
\put(394,62){\circle{16}}
\put(399.5,68.5){\vector(1,1){15}}
\put(399.5,68.5){\vector(1,-1){18}}
\put(390,50){\vector(1,0){28}}
\put(426,50){\circle{16}}
\put(434,50){\vector(3,1){22}}
\put(434,50){\vector(3,-1){22}}
\put(388,43){\vector(1,-1){15}}
\put(385.5,40.5){\vector(2,-3){12}}
\put(382.5,38){\vector(1,-3){6.5}}
\put(410,61){$u$}
\put(405,65){\small 1}
\put(405,40){$u$}
\put(398,40){\small 2}
\put(455,60){$u$}
\put(448,60){\small 1}
\put(455,33){$u$}
\put(447,33){\small 2}
\put(418,82){$\bar d$}
\put(405,82){\small 3}
\put(406,21){$u$}
\put(406,30){\small 4}
\put(398,10){$\bar d$}
\put(399,22){\small 5}
\put(387,7){$\bar d$}
\put(381,10){\small 6}
\put(419,47){\footnotesize $1^{uu}$}
\put(387,59){\footnotesize $1^{u \bar d}$}
\end{picture}

%2al1

\vskip30pt
\begin{picture}(600,100)
%1
\put(0,45){\line(1,0){18}}
\put(0,47){\line(1,0){17.5}}
\put(0,49){\line(1,0){17}}
\put(0,51){\line(1,0){17}}
\put(0,53){\line(1,0){17.5}}
\put(0,55){\line(1,0){18}}
\put(30,50){\circle{25}}
\put(19,46){\line(1,1){15}}
\put(22,41){\line(1,1){17}}
\put(27.5,38.5){\line(1,1){14}}
\put(31,63){\vector(1,1){20}}
\put(37,60){\vector(1,1){20}}
\put(31,38){\vector(1,-1){20}}
\put(37,40){\vector(1,-1){20}}
\put(50,50){\circle{16}}
\put(58,50){\vector(3,2){22}}
\put(58,50){\vector(3,-2){22}}
\put(75,70){$\bar d$}
\put(83,64){\small 1}
\put(75,22){$\bar d$}
\put(83,32){\small 2}
\put(47,86){$u$}
\put(38,82){\small 4}
\put(60,81){$\bar d$}
\put(62,70){\small 3}
\put(39,10){$u$}
\put(35,20){\small 6}
\put(60,13){$u$}
\put(58,27){\small 5}
\put(43,47){\footnotesize $1^{\bar d \bar d}$}
\put(90,48){$=$}
%2
\put(110,45){\line(1,0){18}}
\put(110,47){\line(1,0){17.5}}
\put(110,49){\line(1,0){17}}
\put(110,51){\line(1,0){17}}
\put(110,53){\line(1,0){17.5}}
\put(110,55){\line(1,0){18}}
\put(135,50){\circle{16}}
\put(128,55){\vector(3,2){22}}
\put(128,55){\vector(2,3){15}}
\put(128,45){\vector(3,-2){22}}
\put(128,45){\vector(2,-3){15}}
\put(143,50){\vector(3,1){22}}
\put(143,50){\vector(3,-1){22}}
\put(167,60){$\bar d$}
\put(160,60){\small 1}
\put(167,33){$\bar d$}
\put(160,34){\small 2}
\put(152,62){$\bar d$}
\put(152,74){\small 3}
\put(144,81){$u$}
\put(136,80){\small 4}
\put(152,22){$u$}
\put(151,33){\small 5}
\put(142,12){$u$}
\put(134,16){\small 6}
\put(128,47){\footnotesize $1^{\bar d \bar d}$}
\put(175,48){$+$}
%3
\put(190,48){2}
\put(203,45){\line(1,0){18}}
\put(203,47){\line(1,0){17.5}}
\put(203,49){\line(1,0){17}}
\put(203,51){\line(1,0){17}}
\put(203,53){\line(1,0){17.5}}
\put(203,55){\line(1,0){18}}
\put(233,50){\circle{25}}
\put(222,46){\line(1,1){15}}
\put(225,41){\line(1,1){17}}
\put(230.5,38.5){\line(1,1){14}}
\put(249,62){\circle{16}}
\put(254.5,68.5){\vector(1,1){15}}
\put(254.5,68.5){\vector(1,-1){18}}
\put(245,50){\vector(1,0){28}}
\put(281,50){\circle{16}}
\put(289,50){\vector(3,1){22}}
\put(289,50){\vector(3,-1){22}}
\put(243,43){\vector(1,-1){15}}
\put(240.5,40.5){\vector(2,-3){12}}
\put(237.5,38){\vector(1,-3){6.5}}
\put(265,61){$\bar d$}
\put(260,65){\small 1}
\put(265,37){$\bar d$}
\put(256,40){\small 2}
\put(310,60){$\bar d$}
\put(303,60){\small 1}
\put(310,33){$\bar d$}
\put(302,33){\small 2}
\put(273,82){$\bar d$}
\put(260,82){\small 3}
\put(264,21){$u$}
\put(261,30){\small 4}
\put(253,10){$u$}
\put(254,22){\small 5}
\put(242,7){$u$}
\put(236,10){\small 6}
\put(274,47){\footnotesize $1^{\bar d \bar d}$}
\put(242,59){\footnotesize $1^{\bar d \bar d}$}
\put(320,48){$+$}
%4
\put(335,48){6}
\put(348,45){\line(1,0){18}}
\put(348,47){\line(1,0){17.5}}
\put(348,49){\line(1,0){17}}
\put(348,51){\line(1,0){17}}
\put(348,53){\line(1,0){17.5}}
\put(348,55){\line(1,0){18}}
\put(378,50){\circle{25}}
\put(367,46){\line(1,1){15}}
\put(370,41){\line(1,1){17}}
\put(375.5,38.5){\line(1,1){14}}
\put(394,62){\circle{16}}
\put(399.5,68.5){\vector(1,1){15}}
\put(399.5,68.5){\vector(1,-1){18}}
\put(390,50){\vector(1,0){28}}
\put(426,50){\circle{16}}
\put(434,50){\vector(3,1){22}}
\put(434,50){\vector(3,-1){22}}
\put(388,43){\vector(1,-1){15}}
\put(385.5,40.5){\vector(2,-3){12}}
\put(382.5,38){\vector(1,-3){6.5}}
\put(410,61){$\bar d$}
\put(405,65){\small 1}
\put(410,37){$\bar d$}
\put(401,40){\small 2}
\put(455,60){$\bar d$}
\put(448,60){\small 1}
\put(455,33){$\bar d$}
\put(447,33){\small 2}
\put(418,82){$u$}
\put(405,82){\small 3}
\put(406,18){$\bar d$}
\put(406,30){\small 4}
\put(398,10){$u$}
\put(399,22){\small 5}
\put(387,7){$u$}
\put(381,10){\small 6}
\put(419,47){\footnotesize $1^{\bar d \bar d}$}
\put(387,59){\footnotesize $1^{u \bar d}$}
\end{picture}

%3al1

\vskip30pt
\begin{picture}(600,100)
%1
\put(0,45){\line(1,0){18}}
\put(0,47){\line(1,0){17.5}}
\put(0,49){\line(1,0){17}}
\put(0,51){\line(1,0){17}}
\put(0,53){\line(1,0){17.5}}
\put(0,55){\line(1,0){18}}
\put(30,50){\circle{25}}
\put(19,46){\line(1,1){15}}
\put(22,41){\line(1,1){17}}
\put(27.5,38.5){\line(1,1){14}}
\put(31,63){\vector(1,1){20}}
\put(37,60){\vector(1,1){20}}
\put(31,38){\vector(1,-1){20}}
\put(37,40){\vector(1,-1){20}}
\put(50,50){\circle{16}}
\put(58,50){\vector(3,2){22}}
\put(58,50){\vector(3,-2){22}}
\put(75,70){$u$}
\put(83,64){\small 1}
\put(75,23){$\bar d$}
\put(83,32){\small 2}
\put(47,86){$u$}
\put(38,82){\small 4}
\put(60,81){$u$}
\put(62,70){\small 3}
\put(39,10){$\bar d$}
\put(35,20){\small 6}
\put(60,13){$\bar d$}
\put(58,27){\small 5}
\put(43,47){\footnotesize $1^{u \bar d}$}
\put(90,48){$=$}
%2
\put(110,45){\line(1,0){18}}
\put(110,47){\line(1,0){17.5}}
\put(110,49){\line(1,0){17}}
\put(110,51){\line(1,0){17}}
\put(110,53){\line(1,0){17.5}}
\put(110,55){\line(1,0){18}}
\put(135,50){\circle{16}}
\put(128,55){\vector(3,2){22}}
\put(128,55){\vector(2,3){15}}
\put(128,45){\vector(3,-2){22}}
\put(128,45){\vector(2,-3){15}}
\put(143,50){\vector(3,1){22}}
\put(143,50){\vector(3,-1){22}}
\put(167,60){$u$}
\put(160,60){\small 1}
\put(167,33){$\bar d$}
\put(160,34){\small 2}
\put(152,64){$u$}
\put(152,72){\small 3}
\put(144,81){$u$}
\put(136,80){\small 4}
\put(152,21){$\bar d$}
\put(151,33){\small 5}
\put(142,12){$\bar d$}
\put(134,16){\small 6}
\put(128,47){\footnotesize $1^{u \bar d}$}
\put(175,48){$+$}
%3
\put(190,48){2}
\put(203,45){\line(1,0){18}}
\put(203,47){\line(1,0){17.5}}
\put(203,49){\line(1,0){17}}
\put(203,51){\line(1,0){17}}
\put(203,53){\line(1,0){17.5}}
\put(203,55){\line(1,0){18}}
\put(233,50){\circle{25}}
\put(222,46){\line(1,1){15}}
\put(225,41){\line(1,1){17}}
\put(230.5,38.5){\line(1,1){14}}
\put(249,62){\circle{16}}
\put(254.5,68.5){\vector(1,1){15}}
\put(254.5,68.5){\vector(1,-1){18}}
\put(245,50){\vector(1,0){28}}
\put(281,50){\circle{16}}
\put(289,50){\vector(3,1){22}}
\put(289,50){\vector(3,-1){22}}
\put(243,43){\vector(1,-1){15}}
\put(240.5,40.5){\vector(2,-3){12}}
\put(237.5,38){\vector(1,-3){6.5}}
\put(265,61){$u$}
\put(260,65){\small 1}
\put(260,38){$\bar d$}
\put(253,40){\small 2}
\put(310,60){$u$}
\put(303,60){\small 1}
\put(310,33){$\bar d$}
\put(302,33){\small 2}
\put(273,82){$u$}
\put(260,82){\small 3}
\put(261,21){$u$}
\put(261,30){\small 4}
\put(253,10){$\bar d$}
\put(254,22){\small 5}
\put(242,7){$\bar d$}
\put(236,10){\small 6}
\put(274,47){\footnotesize $1^{u \bar d}$}
\put(242,59){\footnotesize $1^{uu}$}
\put(320,48){$+$}
%4
\put(335,48){2}
\put(348,45){\line(1,0){18}}
\put(348,47){\line(1,0){17.5}}
\put(348,49){\line(1,0){17}}
\put(348,51){\line(1,0){17}}
\put(348,53){\line(1,0){17.5}}
\put(348,55){\line(1,0){18}}
\put(378,50){\circle{25}}
\put(367,46){\line(1,1){15}}
\put(370,41){\line(1,1){17}}
\put(375.5,38.5){\line(1,1){14}}
\put(394,62){\circle{16}}
\put(399.5,68.5){\vector(1,1){15}}
\put(399.5,68.5){\vector(1,-1){18}}
\put(390,50){\vector(1,0){28}}
\put(426,50){\circle{16}}
\put(434,50){\vector(3,1){22}}
\put(434,50){\vector(3,-1){22}}
\put(388,43){\vector(1,-1){15}}
\put(385.5,40.5){\vector(2,-3){12}}
\put(382.5,38){\vector(1,-3){6.5}}
\put(410,61){$u$}
\put(405,65){\small 1}
\put(405,38){$\bar d$}
\put(398,40){\small 2}
\put(455,60){$u$}
\put(448,60){\small 1}
\put(455,33){$\bar d$}
\put(447,33){\small 2}
\put(418,82){$\bar d$}
\put(405,82){\small 3}
\put(406,21){$u$}
\put(406,30){\small 4}
\put(398,10){$u$}
\put(399,22){\small 5}
\put(387,7){$\bar d$}
\put(381,10){\small 6}
\put(419,47){\footnotesize $1^{u \bar d}$}
\put(387,59){\footnotesize $1^{u \bar d}$}
\end{picture}

\vskip30pt
\begin{picture}(600,100)
%5
\put(90,48){$+$}
\put(105,48){2}
\put(118,45){\line(1,0){18}}
\put(118,47){\line(1,0){17.5}}
\put(118,49){\line(1,0){17}}
\put(118,51){\line(1,0){17}}
\put(118,53){\line(1,0){17.5}}
\put(118,55){\line(1,0){18}}
\put(148,50){\circle{25}}
\put(137,46){\line(1,1){15}}
\put(140,41){\line(1,1){17}}
\put(145.5,38.5){\line(1,1){14}}
\put(164,62){\circle{16}}
\put(169.5,68.5){\vector(1,1){15}}
\put(169.5,68.5){\vector(1,-1){18}}
\put(160,50){\vector(1,0){28}}
\put(196,50){\circle{16}}
\put(204,50){\vector(3,1){22}}
\put(204,50){\vector(3,-1){22}}
\put(158,43){\vector(1,-1){15}}
\put(155.5,40.5){\vector(2,-3){12}}
\put(152.5,38){\vector(1,-3){6.5}}
\put(180,61){$\bar d$}
\put(175,65){\small 1}
\put(175,40){$u$}
\put(168,40){\small 2}
\put(225,60){$u$}
\put(218,60){\small 1}
\put(225,33){$\bar d$}
\put(217,33){\small 2}
\put(188,82){$u$}
\put(175,82){\small 3}
\put(176,21){$u$}
\put(176,30){\small 4}
\put(168,10){$\bar d$}
\put(169,22){\small 5}
\put(157,7){$\bar d$}
\put(151,10){\small 6}
\put(189,47){\footnotesize $1^{u \bar d}$}
\put(157,59){\footnotesize $1^{u \bar d}$}
\put(235,48){$+$}
%6
\put(250,48){2}
\put(263,45){\line(1,0){18}}
\put(263,47){\line(1,0){17.5}}
\put(263,49){\line(1,0){17}}
\put(263,51){\line(1,0){17}}
\put(263,53){\line(1,0){17.5}}
\put(263,55){\line(1,0){18}}
\put(293,50){\circle{25}}
\put(282,46){\line(1,1){15}}
\put(285,41){\line(1,1){17}}
\put(290.5,38.5){\line(1,1){14}}
\put(309,62){\circle{16}}
\put(314.5,68.5){\vector(1,1){15}}
\put(314.5,68.5){\vector(1,-1){18}}
\put(305,50){\vector(1,0){28}}
\put(341,50){\circle{16}}
\put(349,50){\vector(3,1){22}}
\put(349,50){\vector(3,-1){22}}
\put(303,43){\vector(1,-1){15}}
\put(300.5,40.5){\vector(2,-3){12}}
\put(297.5,38){\vector(1,-3){6.5}}
\put(325,61){$\bar d$}
\put(320,65){\small 1}
\put(320,40){$u$}
\put(313,40){\small 2}
\put(370,60){$u$}
\put(363,60){\small 1}
\put(370,33){$\bar d$}
\put(362,33){\small 2}
\put(333,82){$\bar d$}
\put(320,82){\small 3}
\put(321,21){$u$}
\put(321,30){\small 4}
\put(313,10){$u$}
\put(314,22){\small 5}
\put(302,7){$\bar d$}
\put(296,10){\small 6}
\put(334,47){\footnotesize $1^{u \bar d}$}
\put(302,59){\footnotesize $1^{\bar d \bar d}$}
\end{picture}

\vskip30pt
\begin{picture}(600,100)
%7
\put(90,48){$+$}
\put(106,48){4}
\put(120,45){\line(1,0){18}}
\put(120,47){\line(1,0){17.5}}
\put(120,49){\line(1,0){17}}
\put(120,51){\line(1,0){17}}
\put(120,53){\line(1,0){17.5}}
\put(120,55){\line(1,0){18}}
\put(150,50){\circle{25}}
\put(139,46){\line(1,1){15}}
\put(142,41){\line(1,1){17}}
\put(147.5,38.5){\line(1,1){14}}
\put(151,63){\vector(1,1){20}}
\put(151,38){\vector(1,-1){20}}
\put(167.5,60){\circle{16}}
\put(167.5,40){\circle{16}}
\put(172,67){\vector(1,1){18}}
\put(172,33){\vector(1,-1){18}}
\put(172,67){\vector(1,-1){17}}
\put(172,33){\vector(1,1){17}}
\put(197,50){\circle{16}}
\put(205,50){\vector(1,1){18}}
\put(205,50){\vector(1,-1){18}}
\put(185,59){$u$}
\put(179,64){\small 1}
\put(181,31){$\bar d$}
\put(179,45){\small 2}
\put(226,65){$u$}
\put(212,65){\small 1}
\put(226,27){$\bar d$}
\put(212,27){\small 2}
\put(194,85){$u$}
\put(183,85){\small 3}
\put(194,10){$\bar d$}
\put(182,10){\small 4}
\put(173,83){$u$}
\put(160,83){\small 5}
\put(173,10){$\bar d$}
\put(160,12){\small 6}
\put(160.5,57){\footnotesize $1^{uu}$}
\put(160.5,37){\footnotesize $1^{\bar d \bar d}$}
\put(190,47){\footnotesize $1^{u \bar d}$}
\end{picture}
\end{figure}

%al2

%\vskip60pt
\begin{figure}
\begin{picture}(600,100)
%1
\put(0,45){\line(1,0){18}}
\put(0,47){\line(1,0){17.5}}
\put(0,49){\line(1,0){17}}
\put(0,51){\line(1,0){17}}
\put(0,53){\line(1,0){17.5}}
\put(0,55){\line(1,0){18}}
\put(30,50){\circle{25}}
\put(19,46){\line(1,1){15}}
\put(22,41){\line(1,1){17}}
\put(27.5,38.5){\line(1,1){14}}
\put(31,63){\vector(1,1){20}}
\put(31,38){\vector(1,-1){20}}
\put(47.5,60){\circle{16}}
\put(47.5,40){\circle{16}}
\put(55,64){\vector(3,2){18}}
\put(55,36){\vector(3,-2){18}}
\put(55,64){\vector(3,-2){18}}
\put(55,36){\vector(3,2){18}}
\put(78,75){$u$}
\put(63,75){\small 1}
\put(78,53){$u$}
\put(70,58){\small 2}
\put(78,41){$\bar d$}
\put(70,36){\small 3}
\put(78,18){$\bar d$}
\put(63,18){\small 4}
\put(54,80){$u$}
\put(41,80){\small 5}
\put(54,13){$\bar d$}
\put(41,13){\small 6}
\put(40.5,56){\footnotesize $1^{uu}$}
\put(40.5,36){\footnotesize $1^{\bar d \bar d}$}
\put(90,48){$=$}
%2
\put(110,45){\line(1,0){19}}
\put(110,47){\line(1,0){21}}
\put(110,49){\line(1,0){23}}
\put(110,51){\line(1,0){23}}
\put(110,53){\line(1,0){21}}
\put(110,55){\line(1,0){19}}
\put(140,60){\circle{16}}
\put(140,40){\circle{16}}
\put(147.5,64){\vector(3,2){18}}
\put(147.5,36){\vector(3,-2){18}}
\put(147.5,64){\vector(3,-2){18}}
\put(147.5,36){\vector(3,2){18}}
\put(128,55){\vector(1,3){11}}
\put(128,45){\vector(1,-3){11}}
\put(170,75){$u$}
\put(155,75){\small 1}
\put(170,53){$u$}
\put(159,59){\small 2}
\put(170,41){$\bar d$}
\put(159,35){\small 3}
\put(170,18){$\bar d$}
\put(155,18){\small 4}
\put(143,86){$u$}
\put(128,84){\small 5}
\put(143,08){$\bar d$}
\put(128,10){\small 6}
\put(133,57){\footnotesize $1^{uu}$}
\put(133,37){\footnotesize $1^{\bar d \bar d}$}
\put(183,48){$+$}
%3
\put(199,48){4}
\put(212,45){\line(1,0){18}}
\put(212,47){\line(1,0){17.5}}
\put(212,49){\line(1,0){17}}
\put(212,51){\line(1,0){17}}
\put(212,53){\line(1,0){17.5}}
\put(212,55){\line(1,0){18}}
\put(242,50){\circle{25}}
\put(231,46){\line(1,1){15}}
\put(234,41){\line(1,1){17}}
\put(239.5,38.5){\line(1,1){14}}
\put(243,63){\vector(1,1){20}}
\put(243,38){\vector(1,-1){20}}
\put(266,80){$u$}
\put(252,80){\small 5}
\put(266,13){$\bar d$}
\put(252,13){\small 6}
\put(262,50){\circle{16}}
\put(270,50){\vector(3,2){17}}
\put(270,50){\vector(3,-2){17}}
\put(247,61){\vector(1,0){40}}
\put(247,39){\vector(1,0){40}}
\put(295,60){\circle{16}}
\put(295,40){\circle{16}}
\put(303,61){\vector(3,1){20}}
\put(303,61){\vector(3,-1){20}}
\put(303,39){\vector(3,1){20}}
\put(303,39){\vector(3,-1){20}}
\put(328,70){$u$}
\put(313,70){\small 1}
\put(328,53){$u$}
\put(310,51){\small 2}
\put(328,41){$\bar d$}
\put(310,44){\small 3}
\put(328,24){$\bar d$}
\put(313,24){\small 4}
\put(270,66){$u$}
\put(260,65){\small 1}
\put(275,49){\small $u$}
\put(270,54){\small 2}
\put(280,45){\small $\bar d$}
\put(270,40){\small 3}
\put(270,27){$\bar d$}
\put(260,28){\small 4}
\put(255,47){\footnotesize $1^{u \bar d}$}
\put(288,57){\footnotesize $1^{uu}$}
\put(288,37){\footnotesize $1^{\bar d \bar d}$}
\put(341,48){$+$}
%4
\put(355,48){2}
\put(368,45){\line(1,0){18}}
\put(368,47){\line(1,0){17.5}}
\put(368,49){\line(1,0){17}}
\put(368,51){\line(1,0){17}}
\put(368,53){\line(1,0){17.5}}
\put(368,55){\line(1,0){18}}
\put(398,50){\circle{25}}
\put(387,46){\line(1,1){15}}
\put(390,41){\line(1,1){17}}
\put(395.5,38.5){\line(1,1){14}}
\put(414,62){\circle{16}}
\put(419.5,68.5){\vector(1,1){15}}
\put(419.5,68.5){\vector(1,-1){18}}
\put(410,50){\vector(1,0){28}}
\put(446,50){\circle{16}}
\put(454,50){\vector(3,1){22}}
\put(454,50){\vector(3,-1){22}}
\put(430,61){$u$}
\put(425,65){\small 1}
\put(427,41){$u$}
\put(425,52){\small 2}
\put(470,62){$u$}
\put(460,59){\small 1}
\put(470,30){$u$}
\put(460,34){\small 2}
\put(436,85){$u$}
\put(424,83){\small 5}
\put(414,37){\circle{16}}
\put(421,32){\vector(3,-1){20}}
\put(421,32){\vector(2,-3){12}}
\put(399,38){\vector(1,-3){8}}
\put(444,22){$\bar d$}
\put(434,32){\small 3}
\put(435,7){$\bar d$}
\put(425,7){\small 4}
\put(409,7){$\bar d$}
\put(397,9){\small 6}
\put(407,59){\footnotesize $1^{uu}$}
\put(439,47){\footnotesize $1^{uu}$}
\put(407,34){\footnotesize $1^{\bar d \bar d}$}
\end{picture}

\vskip60pt
\begin{picture}(600,60)
%5
\put(90,48){$+$}
\put(107,48){2}
\put(125,45){\line(1,0){18}}
\put(125,47){\line(1,0){17.5}}
\put(125,49){\line(1,0){17}}
\put(125,51){\line(1,0){17}}
\put(125,53){\line(1,0){17.5}}
\put(125,55){\line(1,0){18}}
\put(155,50){\circle{25}}
\put(144,46){\line(1,1){15}}
\put(147,41){\line(1,1){17}}
\put(152.5,38.5){\line(1,1){14}}
\put(171,62){\circle{16}}
\put(176.5,68.5){\vector(1,1){15}}
\put(176.5,68.5){\vector(1,-1){18}}
\put(167,50){\vector(1,0){28}}
\put(203,50){\circle{16}}
\put(211,50){\vector(3,1){22}}
\put(211,50){\vector(3,-1){22}}
\put(187,61){$\bar d$}
\put(182,65){\small 1}
\put(184,38){$\bar d$}
\put(182,52){\small 2}
\put(227,62){$\bar d$}
\put(217,59){\small 1}
\put(227,30){$\bar d$}
\put(217,34){\small 2}
\put(193,85){$\bar d$}
\put(181,83){\small 5}
\put(171,37){\circle{16}}
\put(178,32){\vector(3,-1){20}}
\put(178,32){\vector(2,-3){12}}
\put(156,38){\vector(1,-3){8}}
\put(201,22){$u$}
\put(191,32){\small 3}
\put(192,7){$u$}
\put(182,7){\small 4}
\put(166,7){$u$}
\put(158,9){\small 6}
\put(164,59){\footnotesize $1^{\bar d \bar d}$}
\put(196,47){\footnotesize $1^{\bar d \bar d}$}
\put(164,34){\footnotesize $1^{uu}$}
%6
\put(250,48){$+$}
\put(267,48){4}
\put(285,45){\line(1,0){18}}
\put(285,47){\line(1,0){17.5}}
\put(285,49){\line(1,0){17}}
\put(285,51){\line(1,0){17}}
\put(285,53){\line(1,0){17.5}}
\put(285,55){\line(1,0){18}}
\put(315,50){\circle{25}}
\put(304,46){\line(1,1){15}}
\put(307,41){\line(1,1){17}}
\put(312.5,38.5){\line(1,1){14}}
\put(324,68){\circle{16}}
\put(324,32){\circle{16}}
\put(327,53){\vector(3,1){28}}
\put(327,47){\vector(3,-1){28}}
\put(332,70){\vector(3,2){21}}
\put(332,70){\vector(3,-1){23}}
\put(332,30){\vector(3,1){23}}
\put(332,30){\vector(3,-2){21}}
\put(363,60){\circle{16}}
\put(363,40){\circle{16}}
\put(372,60){\vector(3,2){21}}
\put(372,60){\vector(3,-1){23}}
\put(372,40){\vector(3,1){23}}
\put(372,40){\vector(3,-2){21}}
\put(399,71){$u$}
\put(384,74){\small 1}
\put(400,53){$u$}
\put(390,57){\small 2}
\put(400,41){$\bar d$}
\put(390,38){\small 3}
\put(399,19){$\bar d$}
\put(384,19){\small 4}
\put(343,86){$u$}
\put(333,82){\small 5}
\put(343,6){$\bar d$}
\put(333,11){\small 6}
\put(348,68){$u$}
\put(342,68){\small 1}
\put(342,51){$u$}
\put(336,59){\small 2}
\put(348,42){$\bar d$}
\put(338,45){\small 3}
\put(348,23){$\bar d$}
\put(341,26){\small 4}
\put(317,65){\footnotesize $1^{uu}$}
\put(317,29){\footnotesize $1^{\bar d \bar d}$}
\put(356,57){\footnotesize $1^{uu}$}
\put(356,37){\footnotesize $1^{\bar d \bar d}$}
\end{picture}

\vskip60pt
\begin{picture}(600,60)
%7
\put(90,48){$+$}
\put(107,48){4}
\put(125,45){\line(1,0){18}}
\put(125,47){\line(1,0){17.5}}
\put(125,49){\line(1,0){17}}
\put(125,51){\line(1,0){17}}
\put(125,53){\line(1,0){17.5}}
\put(125,55){\line(1,0){18}}
\put(155,50){\circle{25}}
\put(144,46){\line(1,1){15}}
\put(147,41){\line(1,1){17}}
\put(152.5,38.5){\line(1,1){14}}
\put(164,68){\circle{16}}
\put(164,32){\circle{16}}
\put(175,50){\circle{16}}
\put(183,50){\vector(1,1){12}}
\put(183,50){\vector(1,-1){12}}
\put(171,70){\vector(3,2){21}}
\put(171,70){\vector(3,-1){23}}
\put(171,30){\vector(3,1){23}}
\put(171,30){\vector(3,-2){21}}
\put(203,60){\circle{16}}
\put(203,40){\circle{16}}
\put(212,60){\vector(3,2){21}}
\put(212,60){\vector(3,-1){23}}
\put(212,40){\vector(3,1){23}}
\put(212,40){\vector(3,-2){21}}
\put(239,71){$u$}
\put(227,77){\small 1}
\put(240,54){$u$}
\put(230,57){\small 2}
\put(240,40){$\bar d$}
\put(230,37){\small 3}
\put(239,19){$\bar d$}
\put(227,16){\small 4}
\put(183,86){$u$}
\put(173,81){\small 5}
\put(183,6){$\bar d$}
\put(173,12){\small 6}
\put(188,68){$u$}
\put(182,68){\small 1}
\put(182,56){\small $u$}
\put(187,48){\small 2}
\put(181,37){\small $\bar d$}
\put(192,44){\small 3}
\put(188,24){$\bar d$}
\put(182,25){\small 4}
\put(157,65){\footnotesize $1^{uu}$}
\put(157,29){\footnotesize $1^{\bar d \bar d}$}
\put(168,47){\footnotesize $1^{u \bar d}$}
\put(196,57){\footnotesize $1^{uu}$}
\put(196,37){\footnotesize $1^{\bar d \bar d}$}
\end{picture}
\end{figure}

%\vskip60pt
%\newpage

%al3

%\vskip60pt
\begin{figure}
\begin{picture}(600,100)
%1
\put(0,45){\line(1,0){18}}
\put(0,47){\line(1,0){17.5}}
\put(0,49){\line(1,0){17}}
\put(0,51){\line(1,0){17}}
\put(0,53){\line(1,0){17.5}}
\put(0,55){\line(1,0){18}}
\put(30,50){\circle{25}}
\put(19,46){\line(1,1){15}}
\put(22,41){\line(1,1){17}}
\put(27.5,38.5){\line(1,1){14}}
\put(40,68){\circle{16}}
\put(40,32){\circle{16}}
\put(51,50){\circle{16}}
\put(48,70){\vector(3,2){19}}
\put(48,70){\vector(3,-1){22}}
\put(59,50){\vector(3,2){14}}
\put(59,50){\vector(3,-2){14}}
\put(48,30){\vector(3,1){22}}
\put(48,30){\vector(3,-2){19}}
\put(70,80){$u$}
\put(60,84){\small 1}
\put(73,63){$u$}
\put(63,67){\small 2}
\put(75,53){$u$}
\put(61,56){\small 3}
\put(75,41){$\bar d$}
\put(61,39){\small 4}
\put(73,30){$\bar d$}
\put(63,27){\small 5}
\put(70,13){$\bar d$}
\put(60,9){\small 6}
\put(33,65){\footnotesize $1^{uu}$}
\put(33,29){\footnotesize $1^{\bar d \bar d}$}
\put(44,47){\footnotesize $1^{u \bar d}$}
\put(90,48){$=$}
%2
\put(110,45){\line(1,0){19}}
\put(110,47){\line(1,0){21}}
\put(110,49){\line(1,0){23}}
\put(110,51){\line(1,0){23}}
\put(110,53){\line(1,0){21}}
\put(110,55){\line(1,0){19}}
\put(133,65){\circle{16}}
\put(142,50){\circle{16}}
\put(133,35){\circle{16}}
\put(141,68){\vector(1,1){13}}
\put(141,68){\vector(3,-1){18}}
\put(150,50){\vector(3,2){16}}
\put(150,50){\vector(3,-2){16}}
\put(141,32){\vector(3,1){18}}
\put(141,32){\vector(1,-1){13}}
\put(158,80){$u$}
\put(143,79){\small 1}
\put(160,65){$u$}
\put(150,67){\small 2}
\put(170,55){$u$}
\put(158,48){\small 3}
\put(170,39){$\bar d$}
\put(163,44){\small 4}
\put(160,28){$\bar d$}
\put(150,27){\small 5}
\put(158,10){$\bar d$}
\put(143,13){\small 6}
\put(126,62){\footnotesize $1^{uu}$}
\put(135,47){\footnotesize $1^{u \bar d}$}
\put(126,32){\footnotesize $1^{\bar d \bar d}$}
\put(183,48){$+$}
%3
\put(198,48){2}
\put(212,45){\line(1,0){18}}
\put(212,47){\line(1,0){17.5}}
\put(212,49){\line(1,0){17}}
\put(212,51){\line(1,0){17}}
\put(212,53){\line(1,0){17.5}}
\put(212,55){\line(1,0){18}}
\put(242,50){\circle{25}}
\put(231,46){\line(1,1){15}}
\put(234,41){\line(1,1){17}}
\put(239.5,38.5){\line(1,1){14}}
\put(242,30){\circle{16}}
\put(242,22){\vector(2,-3){9}}
\put(242,22){\vector(-2,-3){9}}
\put(262,50){\circle{16}}
\put(270,50){\vector(3,2){17}}
\put(270,50){\vector(3,-2){17}}
\put(247,61){\vector(1,0){40}}
\put(247,39){\vector(1,0){40}}
\put(295,60){\circle{16}}
\put(295,40){\circle{16}}
\put(303,61){\vector(3,1){20}}
\put(303,61){\vector(3,-1){20}}
\put(303,39){\vector(3,1){20}}
\put(303,39){\vector(3,-1){20}}
\put(328,70){$u$}
\put(313,70){\small 1}
\put(328,53){$u$}
\put(307,51){\small 2}
\put(328,41){$u$}
\put(313,46){\small 3}
\put(328,24){$\bar d$}
\put(313,24){\small 4}
\put(224,10){$\bar d$}
\put(234,1){\small 5}
\put(255,10){$\bar d$}
\put(245,1){\small 6}
\put(270,66){$u$}
\put(260,65){\small 1}
\put(280,52){\small $u$}
\put(270,54){\small 2}
\put(280,44){\small $u$}
\put(270,41){\small 3}
\put(270,26){$\bar d$}
\put(260,29){\small 4}
\put(235,27){\footnotesize $1^{\bar d \bar d}$}
\put(255,47){\footnotesize $1^{uu}$}
\put(288,57){\footnotesize $1^{uu}$}
\put(288,37){\footnotesize $1^{u \bar d}$}
%4
\put(341,48){$+$}
\put(356,48){2}
\put(370,45){\line(1,0){18}}
\put(370,47){\line(1,0){17.5}}
\put(370,49){\line(1,0){17}}
\put(370,51){\line(1,0){17}}
\put(370,53){\line(1,0){17.5}}
\put(370,55){\line(1,0){18}}
\put(400,50){\circle{25}}
\put(389,46){\line(1,1){15}}
\put(392,41){\line(1,1){17}}
\put(397.5,38.5){\line(1,1){14}}
\put(400,30){\circle{16}}
\put(400,22){\vector(2,-3){9}}
\put(400,22){\vector(-2,-3){9}}
\put(420,50){\circle{16}}
\put(428,50){\vector(3,2){17}}
\put(428,50){\vector(3,-2){17}}
\put(405,61){\vector(1,0){40}}
\put(405,39){\vector(1,0){40}}
\put(453,60){\circle{16}}
\put(453,40){\circle{16}}
\put(461,61){\vector(3,1){20}}
\put(461,61){\vector(3,-1){20}}
\put(461,39){\vector(3,1){20}}
\put(461,39){\vector(3,-1){20}}
\put(486,70){$u$}
\put(471,70){\small 1}
\put(486,53){$u$}
\put(465,51){\small 2}
\put(486,41){$u$}
\put(471,46){\small 3}
\put(486,24){$\bar d$}
\put(471,24){\small 4}
\put(382,10){$\bar d$}
\put(392,1){\small 5}
\put(413,10){$\bar d$}
\put(403,1){\small 6}
\put(428,66){$u$}
\put(418,65){\small 1}
\put(435,50){\small $u$}
\put(428,54){\small 2}
\put(440,44){\small $\bar d$}
\put(428,41){\small 3}
\put(428,29){$u$}
\put(418,29){\small 4}
\put(393,27){\footnotesize $1^{\bar d \bar d}$}
\put(413,47){\footnotesize $1^{u \bar d}$}
\put(446,57){\footnotesize $1^{uu}$}
\put(446,37){\footnotesize $1^{u \bar d}$}
\end{picture}

\vskip60pt
\begin{picture}(600,60)
%5
\put(90,48){$+$}
\put(105,48){2}
\put(119,45){\line(1,0){18}}
\put(119,47){\line(1,0){17.5}}
\put(119,49){\line(1,0){17}}
\put(119,51){\line(1,0){17}}
\put(119,53){\line(1,0){17.5}}
\put(119,55){\line(1,0){18}}
\put(149,50){\circle{25}}
\put(138,46){\line(1,1){15}}
\put(141,41){\line(1,1){17}}
\put(146.5,38.5){\line(1,1){14}}
\put(149,30){\circle{16}}
\put(149,22){\vector(2,-3){9}}
\put(149,22){\vector(-2,-3){9}}
\put(169,50){\circle{16}}
\put(177,50){\vector(3,2){17}}
\put(177,50){\vector(3,-2){17}}
\put(154,61){\vector(1,0){40}}
\put(154,39){\vector(1,0){40}}
\put(202,60){\circle{16}}
\put(202,40){\circle{16}}
\put(210,61){\vector(3,1){20}}
\put(210,61){\vector(3,-1){20}}
\put(210,39){\vector(3,1){20}}
\put(210,39){\vector(3,-1){20}}
\put(235,70){$\bar d$}
\put(220,70){\small 1}
\put(235,53){$\bar d$}
\put(214,51){\small 2}
\put(235,41){$u$}
\put(220,46){\small 3}
\put(235,24){$\bar d$}
\put(220,24){\small 4}
\put(131,10){$u$}
\put(141,1){\small 5}
\put(162,10){$u$}
\put(152,1){\small 6}
\put(177,65){$\bar d$}
\put(167,65){\small 1}
\put(189,49){\small $\bar d$}
\put(177,54){\small 2}
\put(187,44){\small $u$}
\put(177,41){\small 3}
\put(177,26){$\bar d$}
\put(167,29){\small 4}
\put(142,27){\footnotesize $1^{uu}$}
\put(162,47){\footnotesize $1^{u \bar d}$}
\put(195,57){\footnotesize $1^{\bar d \bar d}$}
\put(195,37){\footnotesize $1^{u \bar d}$}
%6
\put(248,48){$+$}
\put(263,48){2}
\put(277,45){\line(1,0){18}}
\put(277,47){\line(1,0){17.5}}
\put(277,49){\line(1,0){17}}
\put(277,51){\line(1,0){17}}
\put(277,53){\line(1,0){17.5}}
\put(277,55){\line(1,0){18}}
\put(307,50){\circle{25}}
\put(296,46){\line(1,1){15}}
\put(299,41){\line(1,1){17}}
\put(304.5,38.5){\line(1,1){14}}
\put(307,30){\circle{16}}
\put(307,22){\vector(2,-3){9}}
\put(307,22){\vector(-2,-3){9}}
\put(327,50){\circle{16}}
\put(335,50){\vector(3,2){17}}
\put(335,50){\vector(3,-2){17}}
\put(312,61){\vector(1,0){40}}
\put(312,39){\vector(1,0){40}}
\put(360,60){\circle{16}}
\put(360,40){\circle{16}}
\put(368,61){\vector(3,1){20}}
\put(368,61){\vector(3,-1){20}}
\put(368,39){\vector(3,1){20}}
\put(368,39){\vector(3,-1){20}}
\put(393,70){$\bar d$}
\put(378,70){\small 1}
\put(393,53){$\bar d$}
\put(372,51){\small 2}
\put(393,41){$u$}
\put(378,46){\small 3}
\put(393,24){$\bar d$}
\put(378,24){\small 4}
\put(289,10){$u$}
\put(299,1){\small 5}
\put(320,10){$u$}
\put(310,1){\small 6}
\put(335,65){$\bar d$}
\put(325,65){\small 1}
\put(343,48){\small $\bar d$}
\put(335,54){\small 2}
\put(347,44){\small $\bar d$}
\put(335,41){\small 3}
\put(335,29){$u$}
\put(325,29){\small 4}
\put(300,27){\footnotesize $1^{uu}$}
\put(320,47){\footnotesize $1^{\bar d \bar d}$}
\put(353,57){\footnotesize $1^{\bar d \bar d}$}
\put(353,37){\footnotesize $1^{u \bar d}$}
\end{picture}

\vskip60pt
\begin{picture}(600,60)
%7
\put(90,48){$+$}
\put(105,48){4}
\put(119,45){\line(1,0){18}}
\put(119,47){\line(1,0){17.5}}
\put(119,49){\line(1,0){17}}
\put(119,51){\line(1,0){17}}
\put(119,53){\line(1,0){17.5}}
\put(119,55){\line(1,0){18}}
\put(149,50){\circle{25}}
\put(138,46){\line(1,1){15}}
\put(141,41){\line(1,1){17}}
\put(146.5,38.5){\line(1,1){14}}
\put(149,30){\circle{16}}
\put(149,22){\vector(2,-3){9}}
\put(149,22){\vector(-2,-3){9}}
\put(169,50){\circle{16}}
\put(177,50){\vector(3,2){17}}
\put(177,50){\vector(3,-2){17}}
\put(154,61){\vector(1,0){40}}
\put(154,39){\vector(1,0){40}}
\put(202,60){\circle{16}}
\put(202,40){\circle{16}}
\put(210,61){\vector(3,1){20}}
\put(210,61){\vector(3,-1){20}}
\put(210,39){\vector(3,1){20}}
\put(210,39){\vector(3,-1){20}}
\put(235,70){$u$}
\put(220,70){\small 1}
\put(235,53){$u$}
\put(214,51){\small 2}
\put(235,41){$\bar d$}
\put(220,46){\small 3}
\put(235,24){$\bar d$}
\put(220,24){\small 4}
\put(131,10){$u$}
\put(141,1){\small 5}
\put(162,10){$\bar d$}
\put(152,1){\small 6}
\put(177,65){$u$}
\put(167,65){\small 1}
\put(185,50){\small $u$}
\put(177,54){\small 2}
\put(189,44){\small $\bar d$}
\put(177,41){\small 3}
\put(177,26){$\bar d$}
\put(167,29){\small 4}
\put(142,27){\footnotesize $1^{u \bar d}$}
\put(162,47){\footnotesize $1^{u \bar d}$}
\put(195,57){\footnotesize $1^{uu}$}
\put(195,37){\footnotesize $1^{\bar d \bar d}$}
%8
\put(248,48){$+$}
\put(263,48){4}
\put(277,45){\line(1,0){18}}
\put(277,47){\line(1,0){17.5}}
\put(277,49){\line(1,0){17}}
\put(277,51){\line(1,0){17}}
\put(277,53){\line(1,0){17.5}}
\put(277,55){\line(1,0){18}}
\put(307,50){\circle{25}}
\put(296,46){\line(1,1){15}}
\put(299,41){\line(1,1){17}}
\put(304.5,38.5){\line(1,1){14}}
\put(304,62){\vector(2,1){40}}
\put(304,38){\vector(2,-1){40}}
\put(324.5,60){\circle{16}}
\put(324.5,40){\circle{16}}
\put(329,67){\vector(1,1){15}}
\put(329,33){\vector(1,-1){15}}
\put(329,67){\vector(1,-1){17}}
\put(329,33){\vector(1,1){17}}
\put(352,84){\circle{16}}
\put(354,50){\circle{16}}
\put(352,16){\circle{16}}
\put(360,86){\vector(3,2){18}}
\put(360,86){\vector(3,-2){18}}
\put(362,50){\vector(3,2){18}}
\put(362,50){\vector(3,-2){18}}
\put(360,14){\vector(3,2){18}}
\put(360,14){\vector(3,-2){18}}
\put(382,95){$u$}
\put(370,100){\small 1}
\put(382,70){$u$}
\put(372,66){\small 2}
\put(382,58){$u$}
\put(368,58){\small 3}
\put(382,33){$\bar d$}
\put(368,35){\small 4}
\put(382,21){$\bar d$}
\put(372,27){\small 5}
\put(382,-4){$\bar d$}
\put(372,-6){\small 6}
\put(327,80){$u$}
\put(317,77){\small 1}
\put(340,69){$u$}
\put(335,66){\small 2}
\put(340,58){$u$}
\put(333,52){\small 3}
\put(340,35){$\bar d$}
\put(333,42){\small 4}
\put(340,23){$\bar d$}
\put(335,29){\small 5}
\put(327,12){$\bar d$}
\put(317,17){\small 6}
\put(317.5,57){\footnotesize $1^{uu}$}
\put(317.5,37){\footnotesize $1^{\bar d \bar d}$}
\put(345,81){\footnotesize $1^{uu}$}
\put(347,47){\footnotesize $1^{u \bar d}$}
\put(345,13){\footnotesize $1^{\bar d \bar d}$}
\put(10,-25){Fig. 1. Graphic representation of the equations for the six-quark
subamplitudes $A_l$ $(l=1, 2, 3, 4, 5)$}
\put(10,-40){in the case of baryonium $uuu \bar d \bar d \bar d$ $IJ=33$.}
\end{picture}
\end{figure}


\begin{thebibliography}{99}

\bibitem{1}
J.Z. Bai et al, BES Collaboration, Phys. Rev. Lett. {\bf 91}, 022001
(2003).

\bibitem{2}
M. Ablikim et al, BES Collaboration, Phys. Rev. Lett. {\bf 95}, 262001
(2005).

\bibitem{3}
M. Ablikim et al, BESIII Collaboration, arXiv: 1012.3510 [hep-ex].

\bibitem{4}
S.L. Zhu and C.S. Gao, Commun. Theor. Phys. {\bf 46}, 291 (2006).

\bibitem{5}
B. Loiseau and S. Wycech, Int. J. Mod. Phys. A{\bf 20}, 1990 (2005).

\bibitem{6}
C.H. Chang and H.R. Pang, Commun. Theor. Phys. {\bf 43}, 275 (2005).

\bibitem{7}
X.G. He, X.Q. Li and J.P. Ma, Phys. Rev. D{\bf 71}, 014031 (2005).

\bibitem{8}
D.V. Bugg, Phys. Lett. B{\bf 598}, 8 (2004).

\bibitem{9}
I.N. Mishustin, L.M. Satarov, T.J. Burvenich, H. Stoecker and
W. Greiner, Phys. Rev. C{\bf 71},
035201 (2005).

\bibitem{10}
B. Kerbikov, A. Stavinsky and V. Fedotov, Phys. Rev. C{\bf 69}, 055205
(2004).

\bibitem{11}
A. Datta and P.J. O'Donnel, Phys. Lett. B{\bf 567}, 273 (2003).

\bibitem{12}
J.L. Rosner, Phys. Rev. D{\bf 68}, 014004 (2003).

\bibitem{13}
I.J.R. Aitchison, J. Phys. G{\bf 3}, 121 (1977).

\bibitem{14}
J.J. Brehm, Ann. Phys. (N.Y.) {\bf 108}, 454 (1977).

\bibitem{15}
I.J.R. Aitchison and J.J. Brehm, Phys. Rev. D{\bf 17}, 3072 (1978).

\bibitem{16}
I.J.R. Aitchison and J.J. Brehm,
Phys. Rev. D{\bf 20}, 1119, 1131 (1979).

\bibitem{17}
J.J. Brehm, Phys. Rev. D{\bf 21}, 718 (1980).

\bibitem{18}
S.M. Gerasyuta, Yad. Fiz. {\bf 55}, 3030 (1992) [Sov. J. Nucl. Phys.
{\bf 55}, 1693 (1992)].

\bibitem{19}
S.M. Gerasyuta, Nuovo Cimento Soc. Ital. Fis. A{\bf 106}, 37 (1993).

\bibitem{20}
S.M. Gerasyuta, Z. Phys. C{\bf 60}, 683 (1993).

\bibitem{21}
S.M. Gerasyuta and E.E. Matskevich, Phys. Rev. D{\bf 82}, 056002 (2010).

\bibitem{22}
T. Appelqvist and J.D. Bjorken, Phys. Rev. D{\bf 4}, 3726 (1971).

\bibitem{23}
C.C. Chiang, C.B. Chiu, E.C.G. Sudarshan and X. Tata,
Phys. Rev. D{\bf 25}, 1136 (1982).

\bibitem{24}
A. De Rujula, H. Georgi and S.L. Glashow, Phys. Rev. D{\bf 12}, 147
(1975).

\bibitem{25}
V.V. Anisovich, S.M. Gerasyuta and A.V. Sarantsev,
Int. J. Mod. Phys. A{\bf 6}, 625 (1991).

\bibitem{26}
G.'t Hooft, Nucl. Phys. B{\bf 72}, 461 (1974).

\bibitem{27}
G. Veneziano, Nucl. Phys. B{\bf 117}, 519 (1976).

\bibitem{28}
E. Witten, Nucl. Phys. B{\bf 160}, 57 (1979).

\bibitem{29}
O.A. Yakubovsky, Sov. J. Nucl. Phys. {\bf 5}, 1312 (1967).

\bibitem{30}
S.P. Merkuriev and L.D. Faddeev, Quantum Scattering Theory for System
of Few Particles
(Nauka, Moscow 1985) p. 398.

\bibitem{31}
G. Chew, S. Mandelstam, Phys. Rev. {\bf 119}, 467 (1960).

\bibitem{32}
S.M. Gerasyuta and V.I. Kochkin, Phys. Rev. D{\bf 80}, 016006 (2009).

\bibitem{33}
V. Dmitrasinovic, Phys. Rev. D{\bf 67}, 114007 (2003).

\bibitem{34}
F.-Y. Zou, X.-L. Chen and W.-Z. Deng, arXiv: 0710.4365 [hep-ph].

\bibitem{35}
D. Pirjol and C. Schat, Phys. Rev. Lett. {\bf 102}, 152002 (2009).

\bibitem{36}
D. Pirjol and C. Schat, arXiv: 1007.1053 [hep-ph].

\bibitem{37}
D.R. Entem and F. Fernandez, Eur. Phys. J. A{\bf 31}, 649 (2007).

\bibitem{38}
D.R. Entem and F. Fernandez, Phys. Rev. D{\bf 75}, 014004 (2007).

\bibitem{39}
B.S. Zou and H.C. Chiang, Phys. Rev. D{\bf 69}, 034004 (2004).

\bibitem{40}
T. Huang and S.-L. Zhu, Phys. Rev. D{\bf 73}, 014023 (2006).

\end{thebibliography}
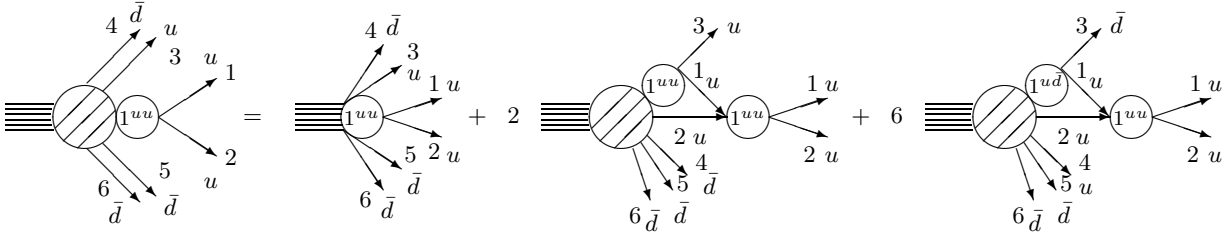
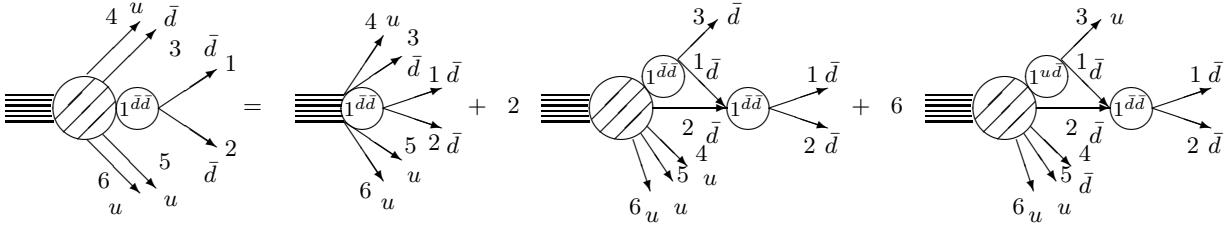
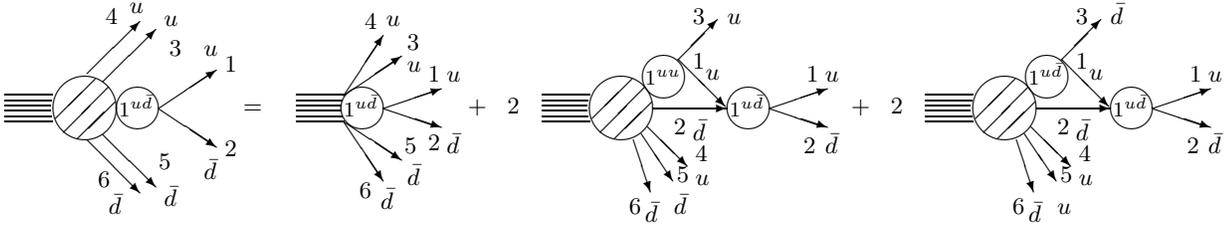
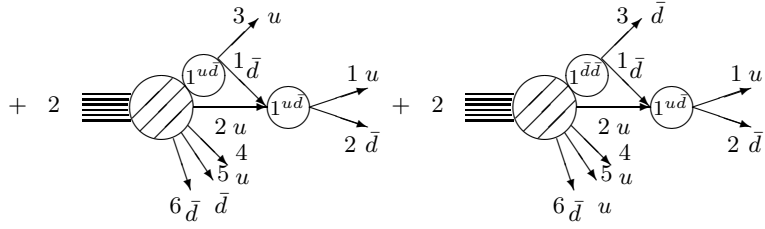
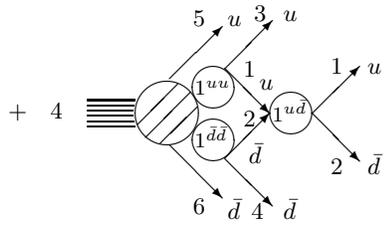
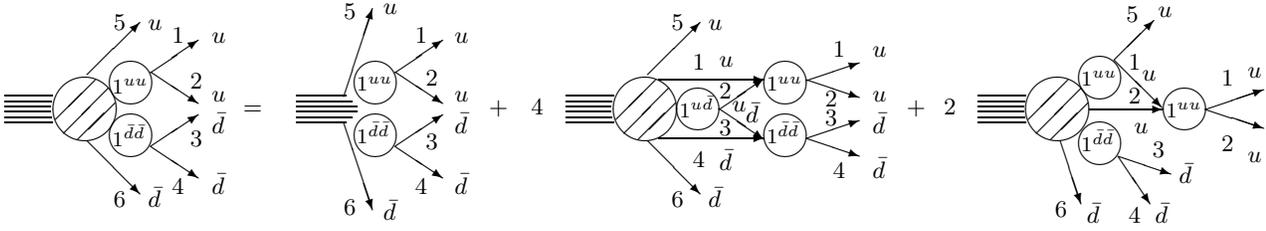
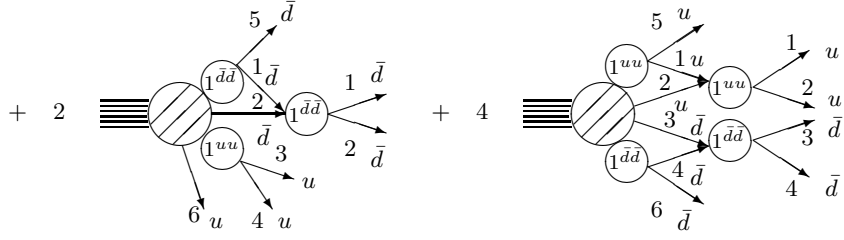
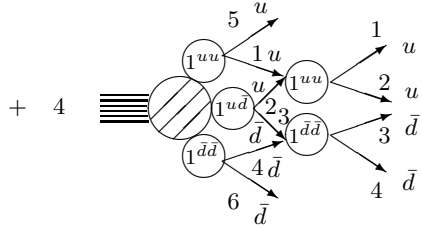
\end{document}